\documentclass[12pt]{iopart}

\usepackage{graphicx,hyperref,iopams,microtype}
\usepackage[square,sort&compress,numbers]{natbib}
\usepackage[varg]{txfonts}
\usepackage{color,journals}

\newcommand{\besselj}[1]{\mathrm{j}_{#1}}
\newcommand{\binom}[2]{{{#1}\choose{#2}}}
\newcommand{\cvector}[1]{\left(\begin{array}{c}#1\end{array}\right)}
\renewcommand{\d}{\mathrm{d}}
\newcommand{\ii}{\mathrm{i}}
\newcommand{\tens}[1]{\boldsymbol{#1}}

\newcommand{\acceptedby}[1]
 {\vspace{28pt plus 10pt minus 18pt}
  \noindent{\small\rm Accepted for publication in: {\it #1}\par}}

\begin{document}

\sloppy

\title
  [Effects of momentum correlations on the cosmic density-fluctuation power spectrum]
  {Kinetic Field Theory: Effects of momentum correlations on the cosmic density-fluctuation power spectrum}
\author
 {Matthias Bartelmann, Felix Fabis, Elena Kozlikin, Robert Lilow, Johannes Dombrowski, Julius Mildenberger}
\address
 {Universit\"at Heidelberg, Zentrum f\"ur Astronomie, Institut f\"ur Theoretische Astrophysik, Philosophenweg 12, 69120 Heidelberg, Germany}

\begin{abstract}
In earlier work, we have developed a Kinetic Field Theory (KFT) for cosmological structure formation and showed that the non-linear density-fluctuation power spectrum known from numerical simulations can be reproduced quite well even if particle interactions are taken into account to first order only. Besides approximating gravitational interactions, we had to truncate the initial correlation hierarchy of particle momenta at the second order. Here, we substantially simplify KFT. We show that its central object, the free generating functional, can be factorized, taking the full hierarchy of momentum correlations into account. The factors appearing in the generating functional, which we identify as non-linearly evolved density-fluctuation power spectra, have a universal form and can thus be tabulated for fast access in perturbation schemes.

In this paper, we focus on a complete evaluation of the free generating functional of KFT, not including particle interactions yet. This implies that the non-linearly evolved power spectra contain a damping term which reflects that structures are being wiped out at late times by free streaming. Once particle interactions will be taken into account, they will compensate this damping. If we suppress this damping in a way suggested by the fluctuation-dissipation relations of KFT, our results show that the complete hierarchy of initial momentum correlations is responsible for a large part of the characteristic non-linear deformation and the mode transport in the density-fluctuation power spectrum. Without any adjustable parameters, KFT accurately reproduces the scale at which non-linear evolution sets in.

Finally, we further develop perturbation theory based on the factorization of the generating functional and propose a diagrammatic scheme for the perturbation terms. 
\end{abstract}



\acceptedby{\NJP}

\section{Introduction}
\label{sec:1}

Based on the pioneering work \cite{2013JSP...152..159D, 2012JSP...149..643D, 2011PhRvE..83d1125M, 2010PhRvE..81f1102M} and in the spirit of \cite{1973PhRvA...8..423M}, we have developed a new theory of structure formation in ensembles of classical particles assumed to be subject to Hamiltonian dynamics and initially correlated in phase space \cite{2016NJPh...18d3020B}. Structurally, the theory resembles a non-equilibrium quantum field theory. Its central object is a free generating functional describing how the initial phase-space distribution of the particles is transported forward in time. The symplectic structure of the Hamiltonian equations and the deterministic trajectories of the classical particles allow substantial simplifications compared to a quantum field theory. Particle interactions are taken into account by applying to the free generating functional an interaction operator which can be expanded into a power series reflecting increasing orders of the interaction. Cumulants of collective fields, such as the macroscopic mass density, can be read off the generating functional by repeated suitable functional derivatives.

Compared to other approaches \cite[see][for a a review and a small selection of recent papers]{1995ApJ...455....7M, 2001A&A...379....8V, 2002PhR...367....1B, 2008JCAP...10..036P, 2011JCAP...06..015A, 2012JCAP...12..013A, 2012JCAP...01..019P, 2014PhRvD..89d3521H}, KFT has several conceptual and methodical advantages. Most importantly, since KFT is based on the Hamiltonian flow in phase space, the problem of shell crossing does not occur. Instead, the flow in phase space is diffeomorphic and, due to the symplectic structure of the Hamiltonian equations, even volume-conserving. Difficulties with multiple streams, as they plague standard perturbation theories based on either the Boltzmann equation or the hydrodynamic equations which assume the existence of uniquely valued velocity fields, are thus avoided by construction in KFT because phase-space trajectories do not cross. Related difficulties caused in Lagrangian perturbation theories \cite[see][for examples]{1970A&A.....5...84Z, 1995A&A...296..575B, 1997GReGr..29..733E, 2013PhRvD..87h3522V, 2012JCAP...06..021R} by functional determinants developing singularities in convergent flows are absent from KFT because the functional determinant of the phase-space flow is constant at unity. Since KFT neither assumes the existence of smooth density or velocity fields nor their uniqueness and avoids taking moments over momentum space, it contains in principle the complete hierarchy of moments and of the particle correlations in configuration and momentum space. As we have shown in \cite{2015PhRvE..91f2120V}, the order of the perturbative approach to particle interactions in KFT controls the order of correlations taken into account in a related BBGKY hierarchy. Since no smooth and uniquely-valued velocity field is assumed to exist, the particle motions in phase space also trace the formation of vorticity on small scales and late times. Finally, the linearity of Hamilton's equations guarantees the existence of a Green's function. Since splitting the Hamiltonian into parts interpreted as unperturbed and perturbed contributions is to a large degree arbitrary, the Green's function can be chosen such that the interaction Hamiltonian becomes small. The latter is one of the main reasons for first-order perturbation theory to be highly successful in KFT, as shown in \cite{2016NJPh...18d3020B}.

We note that the fundamental mathematical framework of KFT is adopted from statistical non-equilibrium field theory and thus routinely used in wide areas of theoretical physics \cite[see][for a recent textbook]{2013LNP...864.....W}. Even though it is not common in cosmology yet, we emphasize that it is not the formalism of KFT that is new, but the application of this formalism to cosmological structure formation, using the particles' phase-space trajectories as fundamental fields. The possibly unfamiliar, but otherwise well established mathematical approach is outweighed by far by the substantial conceptual and methodical advantages of KFT listed above.

Applying this theory to dark-matter structures in cosmology, we showed that even at first order in the interactions between particles on Zel'dovich-type trajectories, the non-linear evolution of the cosmic density-fluctuation power spectrum known from numerical simulations is reproduced very well to redshift zero and to arbitrary wave numbers \cite{2016NJPh...18d3020B}. In addition to the linearization of the interaction operator, we further approximated the initial hierarchy of momentum correlations to second order, i.e.\ we truncated this originally exponential hierarchy after the quadratic momentum-correlation terms. This approach caused our formalism to become quite cumbersome.

Here, we show how the free generating functional of our theory can be fully factorized, taking the complete hierarchy of the initial momentum correlations into account. Besides being more accurate, this entails several major advantages: the formalism simplifies considerably, the development of perturbation theory becomes much more tractable, and the factors have a universal form that can be evaluated and tabulated for fast access in automated evaluations of perturbation terms.

In a first cosmological application of the free generating functional of KFT, we show that including the complete hierarchy of momentum correlations accurately reproduces the scale below which the characteristic non-linear deformations of the density-fluctuation power spectrum set in, that and under which circumstances it leads to mode transport from large to small scales, and that a good fraction of the amplitude of the fully non-linearly evolved power spectrum can already be recovered even in the free theory.

In Sect.~\ref{sec:2}, we briefly summarize the theory, focussing on its free generating functional, specify the initial momentum correlations, and show how the free generating functional can be factorized. First consequences of the theory for cosmological structure formation are described in Sect.~\ref{sec:3}. A systematic perturbative approach to the theory is developed in Sect.~\ref{sec:4} together with a diagrammatic representation of the perturbation terms. We summarize our results in Sect.~\ref{sec:5}. \ref{sec:A1} contains further detail on the correlation function of the initial velocity potential. In \ref{sec:A2}, we present the details of the calculation leading to the factorization of the generating functional.

\section{Generating functional, cumulants, and momentum correlations}
\label{sec:2}

We give a brief overview of our non-equilibrium kinetic theory for correlated classical particle ensembles here as we have developed it in \cite{2016NJPh...18d3020B}. For further detail, we refer the reader to that paper.

\subsection{Generating functional and cumulants}

The central object of the theory is the free generating functional $Z_0[\tens J,\tens K]$ with generator fields $\tens J$ and $\tens K$ coupled to the phase-space coordinates $\tens x = (\tens q, \tens p)$ and the one-particle response fields, respectively. The bold-faced symbols denote tensors bundling contributions from all $N$ particles in the ensemble. Let $\vec x_j = (\vec q_j, \vec p_j)$ be the phase-space coordinates of particle $j$, and $\vec e_j$ an $N$-dimensional unit vector with components $(\vec e_j)_k = \delta_{jk}$, then
\begin{equation}
  \tens x := \vec x_j\otimes\vec e_j\;.
\label{eq:1}
\end{equation}
We define a scalar product between two such tensors by
\begin{equation}
  \langle\tens a,\tens b\rangle =
  \langle\vec a_j\otimes\vec e_j, \vec b_k\otimes\vec e_k\rangle =
  \vec a_j\cdot\vec b_j\;,
\label{eq:2}
\end{equation}
with Einstein's summation convention implied.

This generating functional is the integral
\begin{equation}
  Z_0[\tens J,\tens K] = \int\d\Gamma\,\exp\left(
    \ii\int_0^\infty\d t\left\langle\tens J,\bar{\tens x}\right\rangle
  \right)
\label{eq:3}
\end{equation} 
over the $N$-particle phase space at the initial time $t_0 = 0$. In terms of the Green's function $\mathcal{G}$ of the free equations of motion, the particle trajectories $\bar{\tens x}$ in phase space are
\begin{equation}
  \bar{\tens x}(t) = \mathcal{G}(t,0)\tens x^\mathrm{(i)}-
  \int_0^t\d t'\,\mathcal{G}(t,t')\tens K(t')\;.
\label{eq:4}
\end{equation}
The phase-space measure in (\ref{eq:3}) is
\begin{equation}
  \d\Gamma = P(\tens q^\mathrm{(i)},\tens p^\mathrm{(i)})\,
  \d\tens q^\mathrm{(i)}\d\tens p^\mathrm{(i)}
\label{eq:5}
\end{equation}
with a suitable probability distribution $P$ to be adapted to the initial conditions at hand. For $N$ particles initially correlated in phase space, the probability distribution $P(\tens q^\mathrm{(i)}, \tens p^\mathrm{(i)})$ will be given in (\ref{eq:21}) below.

More explicitly, the $N$-particle Green's function $\mathcal{G}$ is the tensor product
\begin{equation}
  \mathcal{G}(t,t') = G(t,t')\otimes\mathcal{I}_N
\label{eq:6}
\end{equation}
of the matrix-valued, one-particle Green's function
\begin{equation}
  G(t,t') = \left(
    \begin{array}{cc}
      g_{qq}(t,t') & g_{qp}(t,t') \\
      0 & g_{pp}(t,t')
    \end{array}\right)
\label{eq:7}
\end{equation}
and the unit matrix $\mathcal{I}_N$ in $N$ dimensions.

Particle interactions are included by applying an interaction operator to the free generating functional, producing the full generating functional
\begin{equation}
  Z[\tens J,\tens K] = \exp\left(\ii\hat S_\mathrm{I}\right)\,Z_0[\tens J,\tens K]\;,
\label{eq:8}
\end{equation}
with
\begin{equation}
  \hat S_\mathrm{I} = -\int\d 1\,\hat B(-1)v(1)\hat\rho(1)\;.
\label{eq:9}
\end{equation}
Here, $v$ is the interaction potential, and $\hat\rho$ and $\hat B$ are the density- and response-field operators, respectively. The many-particle response field $B$ describes how a particle ensemble responds to a change in the phase-space coordinates of one of its particles. The arguments abbreviate $1 := (t_1,\vec k_1)$ and $-1 := (t_1,-\vec k_1)$. In a Fourier-space representation, these operators are sums over one-particle operators
\begin{eqnarray}
  \hat\rho_j(1) &=&
  \exp\left(-\ii\vec k_1\cdot\frac{\delta}{\ii\delta\vec J_{q_j}(1)}\right)
  \nonumber\\
  \hat B_j(1) &=&
  \left(\ii\vec k_1\cdot\frac{\delta}{\ii\delta\vec K_{p_j}(1)}\right)
  \hat\rho_j(1) =:
  \hat b_j(1)\hat\rho_j(1)
\label{eq:10}
\end{eqnarray}
and the integral in (\ref{eq:9}) is taken over $1 := (t_1, \vec k_1)$. The response-field operator $\hat B$ thus contains a density operator $\hat\rho$.

Correlators of order $n$ in the density, say, are obtained by applying $n$ density operators to $Z[\tens J,\tens K]$,
\begin{equation}
  G_{\rho\ldots\rho}(1\ldots n) = \hat\rho(1)\cdots\hat\rho(n)\,
  Z[\tens J,\tens K]\;.
\label{eq:11}
\end{equation}
As usual in statistical field theory, each of the generator fields $\tens J$ and $\tens K$ is set to zero once all functional derivatives with respect to $\tens J$ or $\tens K$ have been applied.

An approach to a perturbative evaluation of (\ref{eq:11}) begins with expanding the exponential interaction operator $\exp(\ii\hat S_\mathrm{I})$ into a power series, introducing two density operators $\hat\rho$ and one response-field operator $\hat b$ per power of $\hat S_\mathrm{I}$. Thus, for an $n$-th order correlator with $m$-th order particle interaction, we need to evaluate one-particle expressions of the form
\begin{equation}
  \hat\rho_{j_1}(1)\cdots\hat\rho_{j_r}(r)\,
  \left.Z_0[\tens J,\tens K]\right\vert_{\tens J = 0} =
  Z_0[\tens L,\tens K]
\label{eq:12}
\end{equation} 
with $r = n+2m$, and with the particle indices $j_s = 1\ldots N$. The density operators thus replace the generator field $\tens J$ by the shift tensor
\begin{equation}
  \tens L := -\sum_{s=1}^r
  \cvector{\vec k_s\\0}\delta_\mathrm{D}(t-t_s)\otimes\vec e_{j_s}\;.
\label{eq:13}
\end{equation}

We define the position and momentum components of the shift tensor $\tens L$ out by the projections
\begin{eqnarray}
  \vec L_{q_j}(c) &:= \int_0^\infty\d t\,\left\langle
    \tens L(t),G(t,t_c)\cvector{\mathcal{I}_3\\0}\otimes\vec e_j
  \right\rangle\;,\nonumber\\
  \vec L_{p_j}(c) &:= \int_0^\infty\d t\,\left\langle
    \tens L(t),G(t,t_c)\cvector{0\\\mathcal{I}_3}\otimes\vec e_j
  \right\rangle
\label{eq:14}
\end{eqnarray}
and abbreviate
\begin{equation}
  \vec L_{q_j} := \vec L_{q_j}(0)\;,\quad
  \vec L_{p_j} := \vec L_{p_j}(0)
\label{eq:15}
\end{equation}
with $t_0 = 0$. The symbol $\mathcal{I}_3$ denotes the unit matrix in three dimensions.

The subsequent application of a single, one-particle response-field operator $\hat b_{j_c}(c)$ to $Z_0[\tens L,\tens K]$, taken at $\tens K = 0$, simply returns a response-field pre-factor $b_{j_c}(c)$,
\begin{equation}
  \hat b_{j_c}(c)\,\left.Z_0[\tens L,\tens K]\right\vert_{\tens K = 0} =
  b_{j_c}(c)\,Z_0[\tens L,0]\;,
\label{eq:16}
\end{equation}
given by
\begin{equation}
  b_{j_c}(c) = -\ii\vec k_c\cdot\vec L_{p_{j_c}}(c)
\label{eq:17}
\end{equation}
in terms of $\vec L_{p_{j_c}}$ as defined in (\ref{eq:14}). Inserting (\ref{eq:14}) results in
\begin{equation}
  b_{j_c}(c) = \ii\sum_{s=1}^r\left(\vec k_c\cdot\vec k_s\right)\,
  g_{qp}(t_s,t_c)\,\delta_{j_cj_s}\;,
\label{eq:18}
\end{equation} 
which has two important consequences for our later considerations. First, the position-momentum component $g_{qp}(t_s, t_c)$ appears here, evaluated at the two times $t_s$ and $t_c$. Due to causality, the particle ensemble can only respond to causes prior to the response, expressed by $g_{qp}(t_s, t_c) = 0$ for $t_c \ge t_s$. Thus, only response-field factors with $t_c<t_s$ do not vanish. Second, the Kronecker symbol $\delta_{j_cj_s}$ identifies two particle indices. We shall return to these two properties of the response-field factors later.

Our next goal will now be to evaluate the free generating functional $Z_0[\tens L,0]$ after application of an arbitrary number $r$ of density operators. Constructing the phase-space probability distribution $P(\tens q, \tens p)$ in \cite{2016NJPh...18d3020B}, we have assumed that a statistically homogeneous and isotropic, Gaussian random velocity potential $\psi$ exists such that the momentum $\vec p$ at an arbitrary position is its gradient,
\begin{equation}
  \vec p = \vec\nabla\psi\;.
\label{eq:19}
\end{equation}
Continuity then demands that the density contrast $\delta$ is its negative Laplacian,
\begin{equation}
  \delta = -\vec\nabla^2\psi\;.
\label{eq:20}
\end{equation}
Then, the density-fluctuation power spectrum $P_\delta(k)$ specifies the initial phase-space probability distribution $P(\tens q, \tens p)$ completely. As we have shown in \cite{2016NJPh...18d3020B}, it is given by
\begin{equation}
  P(\tens q, \tens p) = \frac{V^{-N}}{\sqrt{(2\pi)^{3N}\det C_{pp}}}\,
  \mathcal{C}(\tens p)\,
  \exp\left(-\frac{1}{2}\tens p^\top C_{pp}^{-1}\tens p\right)\;,
\label{eq:21}
\end{equation}
where $C_{pp}$ is the momentum-correlation matrix to be defined in (\ref{eq:28}) and discussed in the following Section. The correlation operator $\mathcal{C}(\tens p)$ appearing here and defined in \cite{2016NJPh...18d3020B} can safely be approximated by unity,
\begin{equation}
  \mathcal{C}(\tens p) \approx 1\;,
\label{eq:22}
\end{equation}
for correlators evaluated at sufficiently late times if the $q$-$p$-component of the Green's function is unbounded. In the cosmological application we are aiming at here, sufficiently late means that the cosmological scale factor $a$ needs to be much larger than the scale factor $a_\mathrm{i}$ corresponding to the time when the phase-space distribution of the particles is initially set, $a \gg a_\mathrm{i}$. Adopting $a_\mathrm{i} \approx 10^{-3}$ according to the release of the cosmic microwave background, $a > 0.01$ seems safe for approximation (\ref{eq:22}) to hold.

With the probability distribution (\ref{eq:21}), the integrations over the momenta in (\ref{eq:3}) can be carried out straightforwardly. Then, after applying an arbitrary number $r$ of density operators and setting the generator fields to zero, the free generating functional of our microscopic, non-equilibrium, statistical field theory for canonical ensembles of $N$ classical particles enclosed by the volume $V$ can be written in the form
\begin{equation}
  Z_0[\tens L, 0] = V^{-N}\int\d\tens q
  \exp\left(
    -\frac{1}{2}\tens L_p^\top C_{pp}\tens L_p+
    \ii\left\langle\tens L_q,\tens q\right\rangle
  \right)\;,
\label{eq:23}
\end{equation}
valid at all sufficiently late times. We note that this expression for $Z_0[\tens L, 0]$ needs to be summed over all different particle configurations, expressed by the indices $j_1,\ldots,j_r$ appearing in (\ref{eq:12}). The integral in (\ref{eq:23}) is carried out over all particle positions $\tens q$, and $\tens L_q$ and $\tens L_p$ are the position and momentum shift tensors resulting from applying the density operators, with components defined in (\ref{eq:14}).

If only density operators are applied, the shift tensors $\tens L_q$ and $\tens L_p$ will be sums over as many terms as density operators have been applied, with each term representing the contribution of a particle to the density. Since the possible later application of response-field operators will identify two particles per operation, the number of particles involved will be lowered by one for each response-field operator applied. For each particle, one pair of shift vectors $(\vec L_{q_j}, \vec L_{p_j})$ will remain. Thus, if $r$ density operators and $m$ response-field operators have been applied, the number of different particles involved will be $l = r-m = n+m$.

\subsection{Momentum-correlation matrix}

The momentum-correlation matrix is
\begin{equation}
  C_{pp} = \frac{\sigma_1^2}{3}\mathcal{I}_3\otimes\mathcal{I}_N+
  \sum_{j\ne k}C_{p_jp_k}\otimes E_{jk}\;,
\label{eq:24}
\end{equation} 
where the matrix $E_{jk}$ singles out the particles $j$ and $k$ from the ensemble of $N$ particles,
\begin{equation}
  E_{jk} := \vec e_j\otimes\vec e_k\;.
\label{eq:25}
\end{equation}
The amplitude $\sigma_1^2$ is defined as the first moment of the power spectrum $P_\psi$ of the velocity potential $\psi$. Generally, the moments $\sigma_n^2$ are defined by
\begin{equation}
  \sigma_n^2 := \int_kk^{2n}\,P_\psi(k)\;,
\label{eq:26}
\end{equation}
and the velocity-potential power spectrum is related to the density-fluctuation power spectrum $P_\delta$ by
\begin{equation}
  k^4P_\psi(k) = P_\delta(k)
\label{eq:27}
\end{equation} 
due to (\ref{eq:20}).

By definition, the momentum-correlation matrix is given by
\begin{equation}
  C_{p_jp_k} = \int_k\left(\vec k\otimes\vec k\,\right)P_\psi(k)
  \e^{\ii\vec k\cdot\vec q_{jk}}\;,
\label{eq:28}
\end{equation}
where $\vec q_{jk}$ is the separation vector between particles $j$ and $k$. Expression (\ref{eq:28}) is equivalent to
\begin{equation}
  C_{p_jp_k} = -\left(\vec\nabla\otimes\vec\nabla\right)
  \int_kP_\psi(k)\e^{\ii\vec k\cdot\vec q_{jk}} =
  -\left(\vec\nabla\otimes\vec\nabla\right)\,\xi_\psi\left(q_{jk}\right)\;,
\label{eq:29}
\end{equation}
where $\xi_\psi(q)$ is the correlation function of the velocity potential, taken at distance $q$ and normalized to $\sigma_1^2$. Due to isotropy, $\xi_\psi$ can only depend on $q$, but not on the direction of $\vec q$.

Accordingly, the tensor of second derivatives $\vec\nabla\otimes\vec\nabla$ needs to be expressed in terms of derivatives with respect to the particle separation $q$. Let $\hat q$ be the unit vector in the direction of $\vec q$, and by its means define the projectors
\begin{equation}
  \tilde\pi_\parallel := \hat q\otimes\hat q\;,\quad
  \tilde\pi_\perp := \mathcal{I}_3-\tilde\pi_\parallel
\label{eq:30}
\end{equation}
parallel and perpendicular to $\hat q$. Then,
\begin{equation}
  \vec\nabla\otimes\vec\nabla = \tilde\pi_\parallel\,\frac{\d^2}{\d q^2}+
  \tilde\pi_\perp\,q^{-1}\frac{\d}{\d q}
\label{eq:31}
\end{equation}
and
\begin{equation}
  C_{p_jp_k} = -\tilde\pi_\parallel\,\xi_\psi''(q_{jk})-
    \tilde\pi_\perp\,\frac{\xi_\psi'(q_{jk})}{q_{jk}}\;,
\label{eq:32}
\end{equation}
with the prime denoting the derivative with respect to the argument. The correlation function $\xi_\psi$ of the velocity potential and its derivatives $\xi_\psi'$ and $\xi_\psi''$ are worked out in \ref{sec:A1} together with accurate fit formulae convenient for fast numerical evaluations.

The quadratic form
\begin{equation}
  Q := \tens L_p^\top C_{pp}\tens L_p
\label{eq:33}
\end{equation}
remaining in (\ref{eq:23}) splits into two terms by inserting (\ref{eq:24}),
\begin{equation}
  Q = \frac{\sigma_1^2}{3}\sum_j\vec L_{p_j}^{\,2} +
  \sum_{j\ne i}\vec L_{p_i}^\top C_{p_ip_j}\vec L_{p_j}\;.
\label{eq:34}
\end{equation}
Replacing the sum of squares by a squared sum, we can write instead
\begin{equation}
  Q = Q_0-Q_\mathrm{D}+\sum_{j\ne i}\vec L_{p_i}^\top C_{p_ip_j}\vec L_{p_j}
\label{eq:35}
\end{equation}
with the damping terms
\begin{equation}
  Q_0 := \frac{\sigma_1^2}{3}\left(\sum_j\vec L_{p_j}\right)^2\;,\quad
  Q_\mathrm{D} := \frac{\sigma_1^2}{3}
    \sum_{j\ne k}\vec L_{p_j}\cdot\vec L_{p_k}\;.
\label{eq:36}
\end{equation}
We shall see below that $Q_0$ will vanish identically in important cases and that $Q_\mathrm{D}$ has an intuitive and important effect on the time evolution of the density-fluctuation power spectrum. We shall refer to the $Q_0$ and $Q_\mathrm{D}$ as dispersion and diffusion terms, respectively.

\subsection{Factorization of the generating functional}

We now turn to factorizing the generating functional in the form (\ref{eq:23}), which is a lengthy procedure detailed in \ref{sec:A2}. The essential ideas are that, in a statistically homogeneous field, only relative particle coordinates $\vec q_j-\vec q_i$ must matter, and that all these coordinate differences must be statistically indistinguishable.

The final result of the calculations presented in \ref{sec:A2} is that the free generating functional (\ref{eq:23}) for a shift tensor $\tens L$ with contributions from $l$ different particles can be completely factorized,
\begin{equation}
  Z_0[\tens L,0] =
  V^{-l}(2\pi)^3\delta_\mathrm{D}\left(\sum_{j=1}^l\vec L_{q_j}\right)
  \e^{-(Q_0-Q_\mathrm{D})/2}
  \prod_{2\le b<a}^l\int_{k_{ab}}
  \prod_{1\le k<j}^l\left(\Delta_{jk}+\mathcal{P}_{jk}\right)\;.
\label{eq:37}
\end{equation}
The index pairs $(a,b)$ and $(j,k)$ are defined in (\ref{eq:75}) and (\ref{eq:78}), respectively. The function $\mathcal{P}_{jk}$ appearing in each of the factors in (\ref{eq:37}),
\begin{equation}
  \mathcal{P}_{jk}(k_{jk}, \tau) =
  \int_q\,\left\{\e^{g_{qp}^2(\tau,0)\,k_{jk}^2\left(
    a_\parallel\lambda_{jk}^\parallel+a_\perp\lambda_{jk}^\perp
  \right)}-1\right\}
  \e^{\ii\vec k_{jk}\cdot\vec q}\;,
\label{eq:38}
\end{equation}
is a non-linearly (and non-trivially) time-evolved density-fluctuation power spectrum, as we shall discuss in detail in the next Section. The expression $\Delta_{jk}$ abbreviates
\begin{equation}
  \Delta_{jk} :=
  (2\pi)^3\delta_\mathrm{D}\left(\vec k_{jk}\right)\;.
\label{eq:39}
\end{equation}
The wave vectors $\vec k_{jk}$ are defined in (\ref{eq:85}), with the indices $(j,k)$ given in (\ref{eq:75}) and the indices $(a,b)$ defined in (\ref{eq:78}). We shall call the wave vectors $\vec k_{j1}$ \emph{external} because they contain the shift vectors $\vec L_{q_j}$, and the remaining wave vectors $\vec k_{jk}$ with $k\ge2$ \emph{internal} because they can entirely be integrated out. The quantities $\lambda_{jk}^{\parallel, \perp}$ are defined by
\begin{equation}
  \lambda_{jk}^\parallel =
  \frac{\vec L_{p_j}^\top\pi_{jk}^\parallel\vec L_{p_k}}
       {g_{qp}^2(\tau,0)\,k_{jk}^2}
  \quad\mbox{and}\quad
  \lambda_{jk}^\perp =
  \frac{\vec L_{p_j}^\top\pi_{jk}^\perp\vec L_{p_k}}
       {g_{qp}^2(\tau,0)\,k_{jk}^2}
\label{eq:40}
\end{equation}
according to (\ref{eq:100}), with the projectors $\pi_{jk}^\parallel$ and $\pi_{jk}^\perp$ being defined with respect to $\vec k_{jk}$. Let $\hat k$ be the unit vector in the direction of $\vec k_{jk}$, then
\begin{equation}
  \pi_{jk}^\parallel = \hat k\otimes\hat k\;,\quad
  \pi_{jk}^\perp = \mathcal{I}_3-\pi_{jk}^\parallel\;.
\label{eq:41}
\end{equation}
The factorization (\ref{eq:37}) of the free generating functional and the expression (\ref{eq:38}) for the time-evolved density-fluctuation power spectrum are the first main results of our paper. The power spectrum $\mathcal{P}_{jk}$ is particularly relevant for cosmological structure formation.

\section{Cosmological consequences}
\label{sec:3}

For illustrating the cosmological consequences of our results (\ref{eq:37}) and (\ref{eq:38}), we will now reduce (\ref{eq:37}) to the simplest possible case. For $l = 2$, the free generating functional (\ref{eq:37}) contains information on the evolution of the density-fluctuation power spectrum, but neglecting any particle interactions. Thus, this corresponds to the free evolution of the density-fluctuation power spectrum. The free generating functional then shrinks to
\begin{equation}
  Z_0[\tens L, 0] = \frac{(2\pi)^3}{V^2}
  \delta_\mathrm{D}\left(\vec L_{q_1}+\vec L_{q_2}\right)
  \e^{-(Q_0-Q_\mathrm{D})/2}\left(\Delta_{21}+\mathcal{P}_{21}\right)\;.
\label{eq:42}
\end{equation}
The single remaining wave vector is $\vec k_{21} = \vec L_{q_2}$, and the Dirac delta function in (\ref{eq:42}) ensures that $\vec L_{q_2} = -\vec L_{q_1}$. Then, according to (\ref{eq:14}),
\begin{equation}
  \vec L_{p_1} = g_{qp}(t, 0)\vec L_{q_1} = -g_{qp}(t, 0)\vec k_{21}\;,\quad
  \vec L_{p_2} = g_{qp}(t, 0)\vec L_{q_2} =  g_{qp}(t, 0)\vec k_{21}\;,
\label{eq:43}
\end{equation}
and therefore, by the definitions (\ref{eq:40})
\begin{equation}
  \lambda_{21}^\parallel = -1\;,\quad\lambda_{21}^\perp = 0\;.
\label{eq:44}
\end{equation}
This brings (\ref{eq:38}) into the form
\begin{equation}
  \mathcal{P}(k) =
  \int_q\,\left(\e^{-g_{qp}^2(\tau,0)\,k^2a_\parallel}-1\right)
  \e^{-\ii\vec k\cdot\vec q}\;,
\label{eq:45}
\end{equation}
where we have dropped all indices because only the single index pair $21$ now remains to be considered. According to (\ref{eq:111}), this expression turns into
\begin{equation}
  \mathcal{P}(k) \approx g_{qp}^2(\tau,0)\,P_\delta(k)
\label{eq:46}
\end{equation}
in the limit of early times or small wave numbers, which reflects the linear growth of the power spectrum. This emphasizes once more that $\mathcal{P}(k)$ is a density-fluctuation power spectrum, with its time evolution modified by the onset of non-linear evolution.

\begin{figure}[ht]
  \includegraphics[width=\hsize]{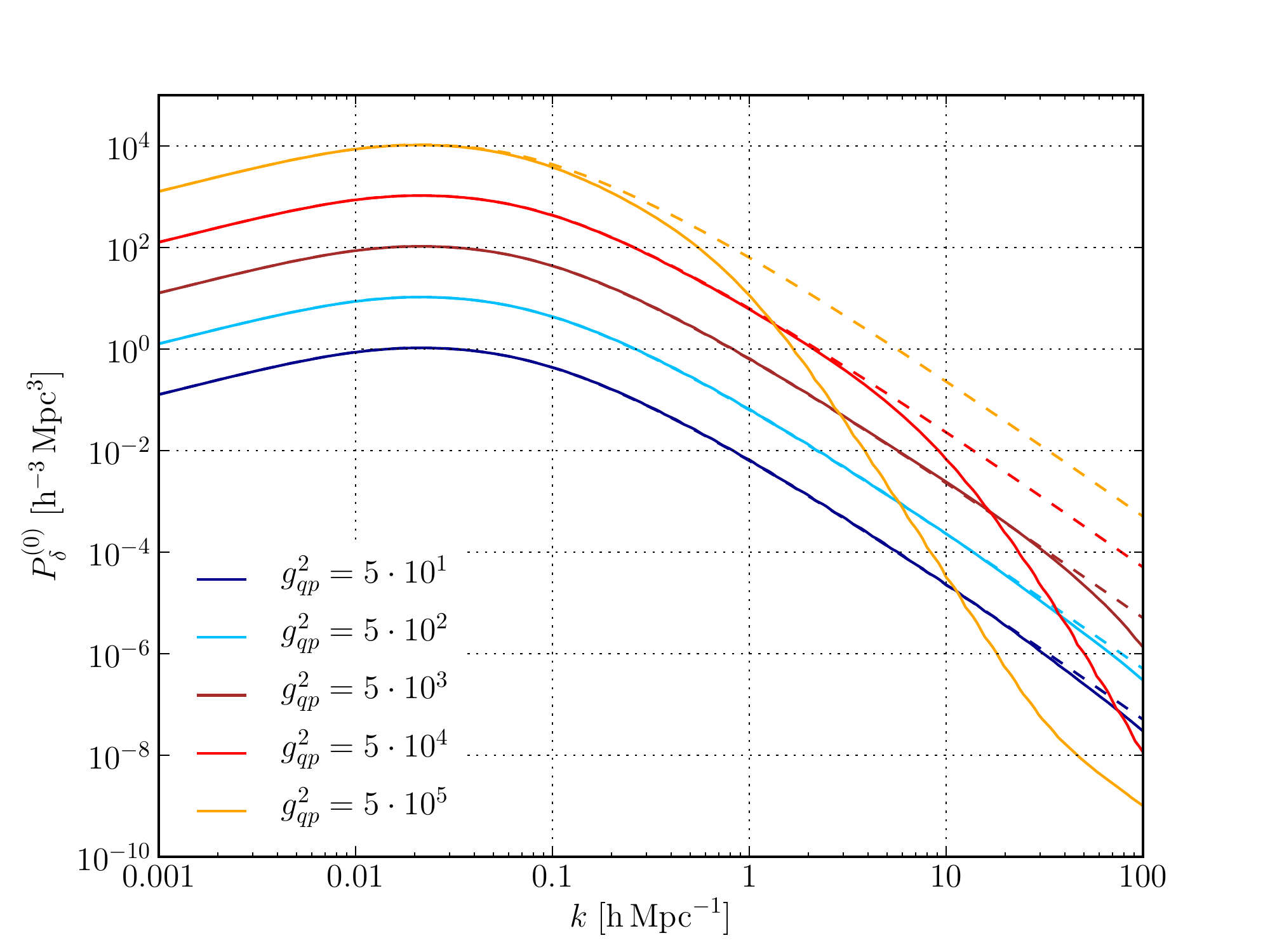}
\caption{The solid curves show the the freely evolved power spectrum $P_\delta^{(0)}$, i.e.\ the non-linearly evolved spectrum $\mathcal{P}$ times the diffusion factor $\exp{Q_D}$, for five different times, expressed by the propagator $g_{qp}(\tau, 0)$. The dashed lines show the initial density-fluctuation power spectrum, linearly evolved to the same times. At large scales, all spectra $P_\delta^{(0)}$ coincide with the linearly evolved spectrum. On smaller scales, diffusion sets in and suppresses the structure, beginning with small-scale structures at early times and progressing towards larger scales as time proceeds.}
\label{fig:1}
\end{figure}

Since the two momentum shift vectors $\vec L_{p_1}$ and $\vec L_{p_2}$ are equal in length and opposite in sign, the dispersion term $Q_0$ from (\ref{eq:36}) vanishes, but the diffusion term $Q_\mathrm{D}$ does not,
\begin{equation}
  Q_\mathrm{D} = -\frac{2\sigma_1^2}{3}g_{qp}^2(\tau, 0)\,k^2\;.
\label{eq:47}
\end{equation}
We show in Fig.~\ref{fig:1} the power spectrum $\mathcal{P}(k)$ from (\ref{eq:45}) times the diffusion factor $\exp(Q_\mathrm{D}/2)$ after different periods of time expressed by the propagator $g_{qp}(\tau, 0)$. This product expresses the free evolution of the power spectrum and will be denoted by $P_\delta^{(0)}$.

Figure~\ref{fig:1} illustrates two aspects of the evolution. First, the non-linearly evolved spectrum $\mathcal{P}$ coincides with the linearly evolving density-fluctuation spectrum for all times at sufficiently small wave numbers, i.e.\ for sufficiently large structures. Since the free generating functional does not contain interactions between individual particles, structures are damped by free streaming, provided they are small enough. Thus, diffusion proceeds from small to larger scales as time progresses. Starting from (\ref{eq:38}) and including the diffusion term, we can write the expression for the damped power spectrum briefly as
\begin{equation}
  \e^{y_0}\mathcal{P} =
  \int_q\e^{y_0}\left(\e^{y}-1\right)\e^{\ii\vec k\cdot\vec q}\;,
\label{eq:48}
\end{equation}
with $y := g_{qp}^2\,k^2\,(a_\parallel\lambda^\parallel+a_\perp\lambda^\perp)$ and $y\to y_0 = Q_\mathrm{D}/2$ for $q\to 0$.

\begin{figure}[ht]
  \includegraphics[width=\hsize]{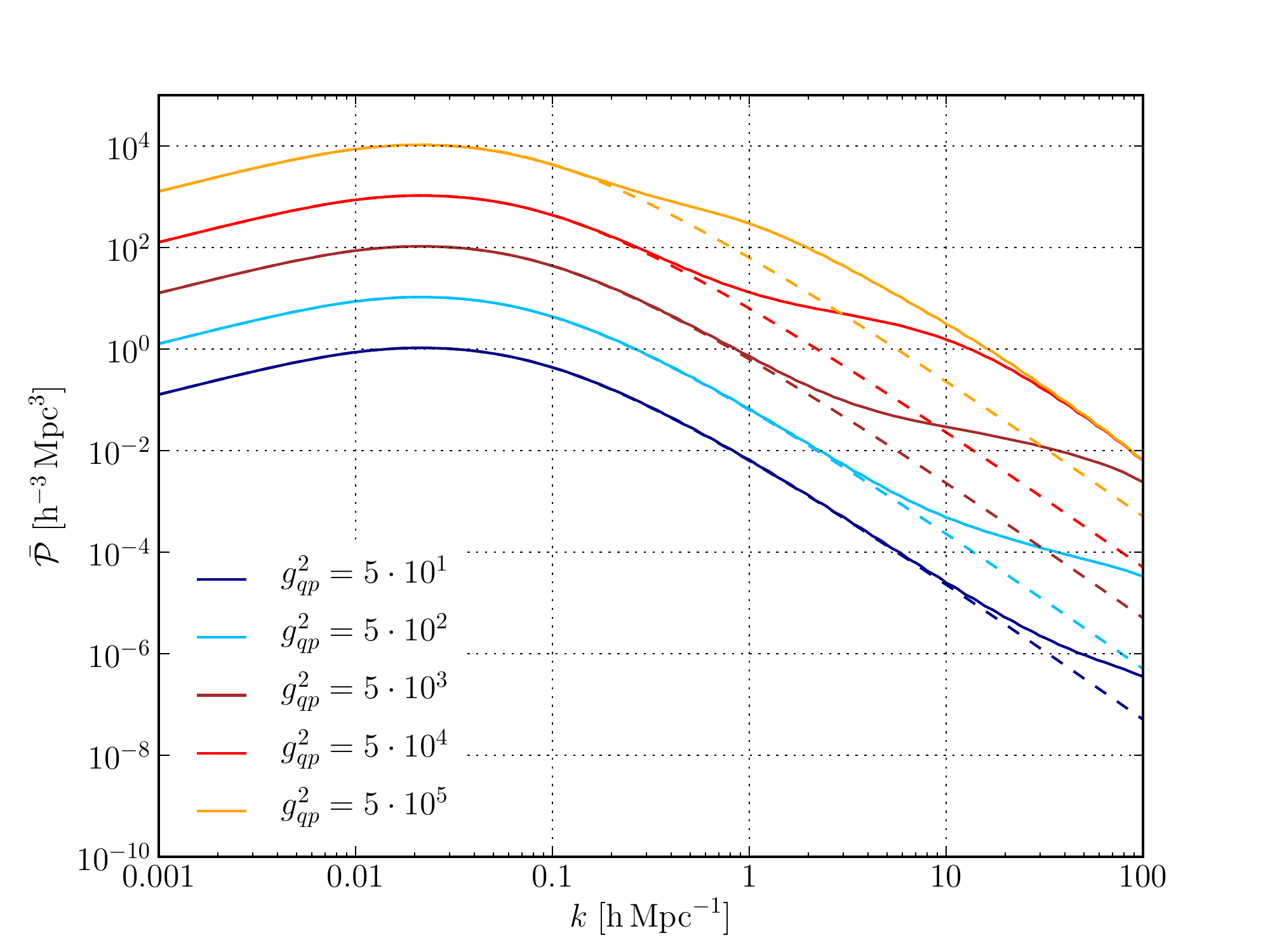}
\caption{Non-linearly evolved power spectra are shown here for the same times as in Fig.~\ref{fig:1}, but now with the late-time diffusion suppressed as specified in (\ref{eq:50}). In this representation undamped by free streaming, the non-linearly evolving power spectra indicate how structure formation proceeds from small to large scales.}
\label{fig:2}
\end{figure}

Our analysis allows us to largely remove this damping effect from the evolution of the power spectrum. We do so by introducing a parameter $0\le\alpha\le1$ meant to gradually switch on the momentum correlation in a thought experiment. We can then write
\begin{eqnarray}
  \e^{y_0}\mathcal{P} &=
  \int_q\e^{y_0}\left(e^y-1\right)\e^{\ii\vec k\cdot\vec q} =
  \e^{y_0}\int_q\int_0^1\d\alpha
  \frac{\partial}{\partial\alpha}\left(\e^{\alpha y}\right)
  \e^{\ii\vec k\cdot\vec q} \nonumber\\ &=
  \e^{y_0}\int_0^1\d\alpha\int_qy\,\e^{\alpha y+\ii\vec k\cdot\vec q} =
  \int_0^1\d\alpha\,\e^{y_0(1-\alpha)}
  \int_qy\,\e^{\alpha(y+y_0)+\ii\vec k\cdot\vec q}\;.
\label{eq:49}
\end{eqnarray} 
We have arranged terms such that one contribution to the damping term builds up as $\alpha$ increases and the correlation strengthens, while the other contribution is strongest for $\alpha = 0$ and decreases as the correlation grows. We ignore this latter term, thus suppressing the contribution to damping present without particle correlations, and replace (\ref{eq:49}) by
\begin{equation}
  \e^{y_0}\mathcal{P} \to \bar\mathcal{P} =
  \int_0^1\d\alpha\,\int_qy\,\e^{\alpha(y+y_0)+\ii\vec k\cdot\vec q} =
  \int_0^1\d\alpha\,\e^{\alpha y_0}\frac{\partial}{\partial\alpha}
  \int_q \e^{\alpha y+\ii\vec k\cdot\vec q}\;.
\label{eq:50}
\end{equation}
If the remaining damping contribution was absent, $\bar\mathcal{P}=\mathcal{P}$. By construction, this operation removes that part of the power suppression due to uncorrelated particle momenta, which dissipate the newly formed structures by free streaming. The resulting power spectra $\bar\mathcal{P}$ are shown in Fig.~\ref{fig:2} for the same times as the power spectra $P_\delta^{(0)}$ in Fig.~\ref{fig:1}.

The undamped power spectra shown in Fig.~\ref{fig:2} illustrate how structure builds up beginning at small and proceding towards larger scales as time progresses. At the latest time used in Figs.~\ref{fig:1} and \ref{fig:2}, approximately corresponding to the present time in the standard $\Lambda$CDM cosmology, non-linear structure formation has reached wave numbers of $k\sim0.2-0.3\,h\,\mathrm{Mpc}^{-1}$, in excellent agreement with the transition to non-linearity revealed by numerical simulations.

We can of course not expect our simple procedure to return the correct amplitude of the power spectrum because we have not included particle interactions yet and thus have not gone beyond the lowest-order approximation of the generating functional, expressed by setting $l = 2$. Nonetheless, the shape of the undamped, non-linearly evolved density-fluctuation power spectrum at late cosmic times, which is obtained in a completely parameter-free way in our theory, reflects the shape of the fully non-linear power spectrum found in numerical simulations very well. We should like to emphasize that the procedure according to (\ref{eq:50}) applied to the non-linearly evolved power spectrum finds a deeper explanation in the fluctuation-dissipation theorems following from our kinetic theory of structure formation. We shall discuss this in detail in a separate paper now in preparation, dedicated to fluctuation-dissipation relations in KFT. The response to fluctuations in the particle density is mediated by particle interactions. This supports the intuition that the excess diffusion seen in the curves of Fig.~\ref{fig:1} will be compensated by particle interactions, which are not included yet in our analysis of the free generating functional. Based on the factorization of the free generating functional demonstrated in this paper, and using the diagrammatic approach to the factorization to be proposed in the following Section, we shall develop a systematic perturbative approach to non-linear structure formation including particle interactions in another dedicated paper.

The fact that the undamped, non-linearly evolved power spectra $\bar\mathcal{P}$ shown in Fig.~\ref{fig:2} already resemble the shape of the non-linearly developed power spectra of fully numerical simulations shows that this shape is not set by the particle-particle interactions, but rather by their only property we have included here, which is the correlation of their initial momenta. The deformation of the power spectrum beginning at $k\gtrsim0.2-0.3$ characteristic for the late-time non-linear evolution of cosmic structures is therefore determined by the statistical initial conditions of the particle ensemble, specifically by the initial correlation properties of the particle momenta, presumably imprinted by the Gaussian random density-fluctuation field generated by cosmological inflation.

We notice also that the curves shown in Fig.~\ref{fig:2} do not diverge for large wave numbers $k$. Rather, they approach an asymptotic slope close to that of the linearly evolved density-fluctuation power spectrum, which tends to $k^{-3}$ for cold dark matter. This indicates that there is no inherent limit to applying our theory to arbitrarily large $k$, which was to be expected because our phase-space approach does by construction not suffer from any problems with multiple streams. Of course, the amplitude of the density-fluctuation power spectrum will be affected by particle interactions. However, the factorization (\ref{eq:37}) of the free generating functional together with the asymptotic behaviour of the curves shown in Fig.~\ref{fig:2} both indicate that non-linearly evolved density-fluctuation power spectra evaluated at an arbitrary order of particle interactions will be a convolution of curves with an asymptotic slope near $k^{-3}$, multiplied by the subsequent application of damping factors. It is plausible that the result will retain the same asymptotic slope, but we will have to demonstrate that. However, the regular asymptotic behaviour of our non-linearly evolved power spectra $\bar\mathcal{P}(k)$ for large $k$ suggests that KFT can in this sense be extended to arbitrarily large wave numbers.

\begin{figure}[ht]
  \includegraphics[width=\hsize]{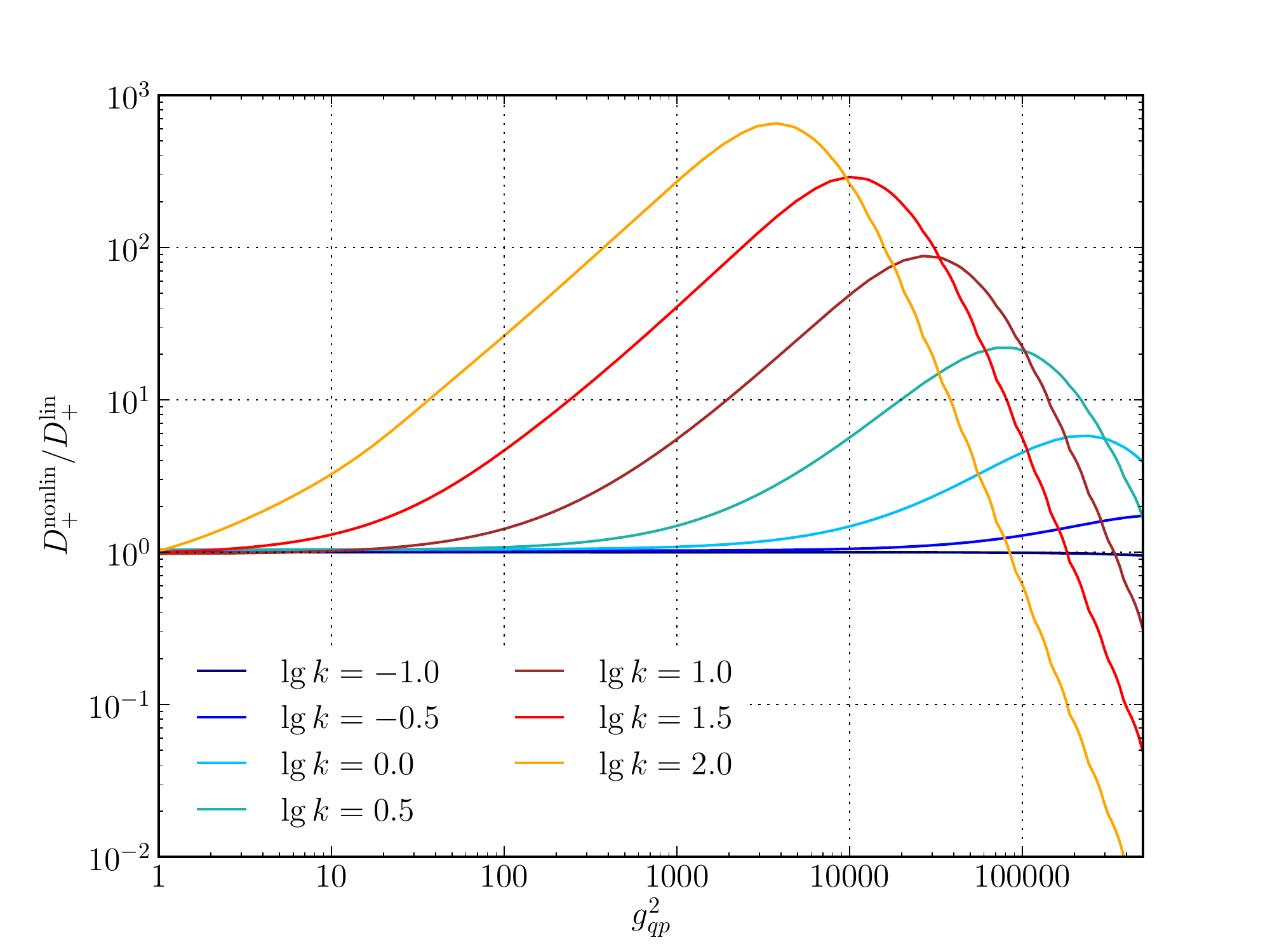}
\caption{This figure shows the derivative of the non-linearly growing power spectrum $\bar\mathcal{P}$ with respect to the squared propagator $g_{qp}^2 = D_+^2$, divided by the initial power spectrum $P_\delta(k)$. If the power spectrum grew linearly, the curves would be flat at unity. The figure shows that, for late times or small-scale structures, the power spectrum grows far beyond the linear growth. The decrease of the curves at late times reflects the remaining damping, which will be compensated by gravitational interaction.}
\label{fig:3}
\end{figure}

Figure~\ref{fig:3} shows the derivative of $\bar\mathcal{P}$ from (\ref{eq:50}) with respect to $g_{qp}^2$, divided by the initial density-fluctuation power spectrum $P_\delta(k)$. As the limit (\ref{eq:46}) shows,
\begin{equation}
  \frac{1}{P_\delta}\frac{\partial}{\partial g_{qp}^2}\mathcal{P} \to 1
\label{eq:51}
\end{equation}
for large scales or early times. Since $g_{qp} = D_+$, i.e.\ the linear growth factor of density perturbations, any deviations of this expression from unity indicates structure growth different from the growth expected in linear theory.

The curves in Fig.~\ref{fig:3} illustrate how the growth factor of cosmic density fluctuations becomes scale dependent. They show that, for small-scale structures or at late times, the power spectrum grows substantially more rapidly than expected from linear theory. The decrease of the curves for small-scale structure at late times indicates the remaining damping, which will be counteracted by gravity once particle interactions will be included.

\begin{figure}[ht]
  \includegraphics[width=\hsize]{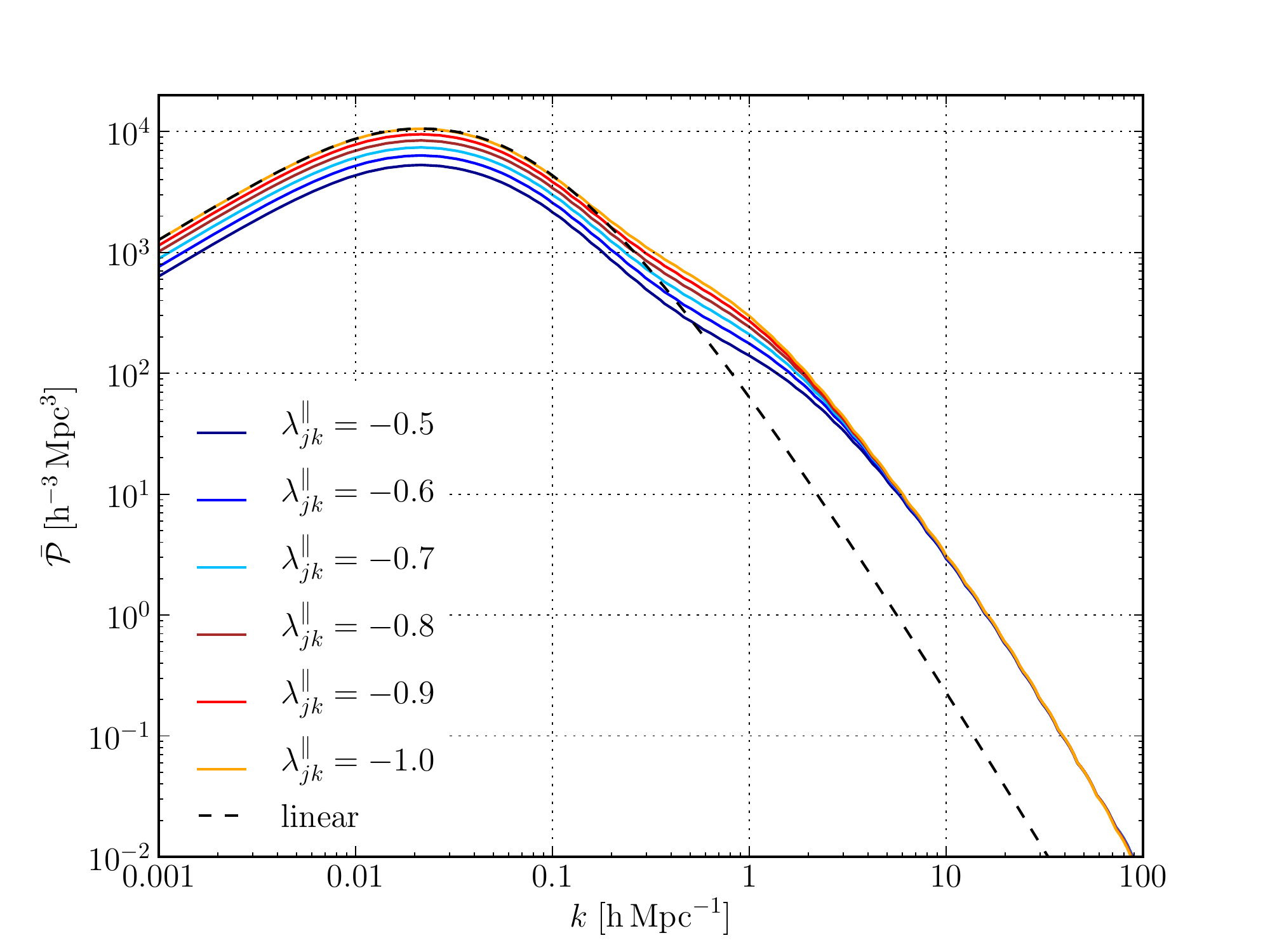}
\caption{Power spectra $\bar\mathcal{P}_{jk}$ for late times and for configurations with $\lambda_{jk}^\parallel\ge-1$, indicating that the momentum shift vectors $\vec L_{p_j}$ and $\vec L_{p_k}$ are misaligned with the separation vector connecting the two particles. In this case, the power at large scales is reduced compared to the linearly evolved power spectrum, and enhanced at small scales. This shows that such a misalignment causes mode transport from large to small scales.}
\label{fig:4}
\end{figure}

So far, we have evaluated the free generating functional for the simplest possible case, i.e.\ $l=2$, where $\lambda_{jk}^\parallel = -1$ and $\lambda_{jk}^\perp = 0$. This happens if the two momentum shift vectors $\vec L_{p_j}$ and $\vec L_{p_k}$ appearing in $\mathcal{P}_{jk}$ are equal and opposite to each other, and if they are aligned with the wave vector $\vec k_{jk}$. If the momentum shift vectors are however not aligned with the wave vector $\vec k_{jk}$, the parallel projection $\lambda^\parallel_{jk}>-1$. This expresses a configuration in which the momenta of two particles are misaligned with the separation vector between the two particles. Power spectra $\bar\mathcal{P}_{jk}$ for such cases are shown in Fig.~\ref{fig:4}.

The figure shows that such a misalignment between particle momenta and their separation vector leads to a reduction of power relative to the linearly evolved power spectrum at large scales, and to an enhancement on small scales. This causes mode transport from large to small scales, adding to the characteristic deformation of the power spectrum by non-linear evolution.

The non-linear growth of the power spectrum and the mode transport contained in $\bar\mathcal{P}_{jk}$ and illustrated by various examples in Figs.~\ref{fig:1} through \ref{fig:4} thus leads to a characteristic deformation of the power spectrum compared to its linearly evolved shape. As our derivation shows, this deformation does not reflect the gravitational interaction between the particles, but is a consequence exclusively of the initial correlations of the particle momenta. Particles with momenta aligned with their separation vector lead to an enhancement of power and thus to structure growth on small and intermediate scales. The more these particle momenta are misaligned with their separation vector, the more power is transported from large to small scales. Even before any gravitational interaction between the particles is taken into account, the full hierarchy of initial momentum correlations gives rise to a characteristic deformation of the power spectrum. Initial momentum correlations of the dark-matter particles thus contribute substantially not only to the amplitude of cosmic structures, but also to characteristic re-distribution of power from larger to smaller scales.

\section{Perturbative and diagrammatic approach}
\label{sec:4}

\subsection{Expansion of the generating functional}

The main goal of this Section is to develop a systematic approach to evaluating (\ref{eq:37}). It begins by realising that the product over the index pair $(j,k)$ in (\ref{eq:37}) can be expanded into a sum ordered by an increasing power of power-spectrum factors. Since there are
\begin{equation}
  F = \frac{l(l-1)}{2}
\label{eq:52}
\end{equation}
factors in this product, we can write formally
\begin{equation}
  \prod_{j>k}\left(\Delta_{jk}+\mathcal{P}_{jk}\right) =
  \sum_{f=0}^F\binom{F}{f}\,\Delta^{F-f}\,\mathcal{P}^f\;.
\label{eq:53}
\end{equation}
If the power-spectrum factors $\mathcal{P}$ are small enough, this sum can be truncated at low powers $f$.

According to (\ref{eq:37}), all internal wave vectors $\vec k_{ab} = \vec k'_{ab}$ are then to be integrated over. For evaluating the result of this sequence of integrals applied to the sum terms in (\ref{eq:53}), it is important to see how many of these integrals can trivially be carried out due to the $\Delta$ factors appearing in these sum terms. For $f = 0$, for example, the remaining integrals set all of the internal wave vectors to zero, leaving $\vec k_{j1} = \vec L_{q_j}$ and
\begin{equation}
  \prod_{a>b}\int_{k_{ab}}\prod_{j>k}\Delta_{jk} =
  \prod_{j=2}^l\Delta\left(\vec L_{q_j}\right)\;.
\label{eq:54}
\end{equation}
This is a pure shot-noise term. Identifying and counting the power-spectrum factors with external or internal wave numbers thus allows to simplify the remaining expressions substantially.

Let us illustrate this simplification with one more example. For $f = 1$, the single power-spectrum factor can depend either on an external or an internal wave number. If it is external, the $\Delta$ factors set all internal wave numbers to zero as for $f = 0$ before, leaving again the external wave numbers $\vec k_{j1} = \vec L_{q_j}$. All of these wave vectors $\vec k_{j1}$ except one are also set to zero by the remaining $\Delta$ factors. Without loss of generality, we can label the one non-zero wave vector by the index $j = 2$ because the particles are indistinguishable. Then, the power-spectrum factor $\mathcal{P}(\vec L_{q_2})$ appears besides the remaining $\Delta$ factors. If, however, the single power-spectrum factor depends on an internal wave number, all except one internal wave numbers are integrated out. Again without loss of generality, we can label the only remaining internal wave number $\vec k_{32} = \vec k'_{32}$. The external wave vectors are then
\begin{equation}
  \vec k_{21} = \vec L_{q_2}+\vec k'_{32}\;,\quad
  \vec k_{31} = \vec L_{q_3}-\vec k'_{32}
\label{eq:55}
\end{equation}
and $\vec k_{j1} = \vec L_{q_j}$ for $j > 3$. All external wave vectors appear as arguments of $\Delta$ factors in this case. The remaining integration over $\vec k'_{32}$ sets $\vec k'_{32} = \vec L_{q_3}$ by means of the factor $\Delta_{31}$ and leaves $\vec k_{21} = \vec L_{q_2}+\vec L_{q_3}$.

There are $l-1$ external and $F-l+1$ internal wave vectors. Thus, there are $l-1$ possibilities for choosing an external and $F-l+1$ possibilities for choosing an internal wave number, allowing us to write
\begin{equation}
\fl
  \prod_{a>b}\int_{k_{ab}}\binom{F}{1}\,\mathcal{P}\Delta^{F-1} =
  \binom{l-1}{1}\,\mathcal{P}_{21}(\vec L_{q_2})\prod_{j=3}^l\Delta_{j1}+
  \binom{F-l+1}{1}\,\mathcal{P}_{32}(\vec L_{q_3})
  \Delta_{21}(\vec L_{q_1})\prod_{j=4}^l\Delta_{j1}\;,
\label{eq:56}
\end{equation}
where the overall Dirac delta distribution
\begin{equation}
  \delta_\mathrm{D}\left(\sum_{j=1}^l\vec L_{q_j}\right)
\label{eq:57}
\end{equation}
in (\ref{eq:37}) allows us to replace $\Delta_{21}(\vec L_{q_2}+\vec L_{q_3})$ by $\Delta_{21}(\vec L_{q_1})$.

From these examples for $f = 0$ and $f = 1$, a straightforward scheme emerges for evaluating the sum terms in (\ref{eq:53}) and the subsequent integrations over all internal wave vectors:
\begin{enumerate}
  \item Expand each term into a sum ordered by the number of power-spectrum factors depending on external wave numbers,
  \begin{equation}
  \label{eq:58}
    \binom{F}{f}\Delta^{F-f}\mathcal{P}^f =
    \sum_{e=0}^f\binom{l-1}{e}\binom{F-l+1}{f-e}\,
    \mathcal{P}_\mathrm{ext}^e\mathcal{P}_\mathrm{int}^{f-e}
    \Delta_\mathrm{ext}^{l-e-1}\Delta_\mathrm{int}^{F-l+1-f+e}\;,
  \end{equation}
  where the subscripts `ext' and `int' indicate that respective factors depend on external or internal wave vectors.
  \item Use the $\Delta$ factors depending on internal wave vectors to set $(F-l+1-f+e)$ of the internal wave vectors to zero. Label the remaining $(f-e)$ internal wave vectors beginning with $\vec k_{32}$ and determine the external wave vectors $\vec k_{j1}$ according to (\ref{eq:85}).
  \item Use the $\Delta$ factors depending on external wave vectors and the integrals over the remaining internal wave vectors to eliminate as many of the internal wave numbers as possible from the external wave vectors $\vec k_{j1}$ and from the arguments of the power-spectrum factors depending on internal wave numbers.
\end{enumerate}
This procedure lends itself to be evaluated by a symbolic computer code returning the terms appearing in the sum (\ref{eq:58}), given $l$ and $f$. We are now in the process of developing such a code, aiming at possibly completely automatizing the evaluation of perturbation terms.

\subsection{Summing over particles}

As often in perturbation theories, diagrams are useful to keep an overview of the terms involved. We shall now develop suitable diagrams for the perturbative approach to our non-equilibrium field theory for correlated classical particle ensembles.

We return to the free generating functional $Z_0[\tens L,0]$, evaluated at a specific shift tensor $\tens L$, with the generator field $\tens K$ set to zero. The position and momentum components of $\tens L$ are given in (\ref{eq:14}) or, after inserting (\ref{eq:13}) there, by
\begin{eqnarray}
  \vec L_{q_j} &= -\sum_{s=1}^r\vec k_s\,\delta_{jj_s}\;,\nonumber\\
  \vec L_{p_j} &= -\sum_{s=1}^r\vec k_s\,g_{qp}(t_s,0)\,\delta_{jj_s}\;.
\label{eq:59}
\end{eqnarray}
In these expressions, the index $1\le j\le l$ identifies the $l$ particles whose positions are being correlated, while the indices $j_s$ assign particles to wave vectors with field labels $s$. The Kronecker symbols appear in (\ref{eq:59}) because only such phase-space positions contribute to the density which are occupied by particles. Given a specific set of particle indices $\{j_1,\ldots,j_r\}$, we write
\begin{equation}
  Z_0[\{j_1,\ldots,j_r\}] := Z_0[\tens L,0]
\label{eq:60}
\end{equation}
to denote the particle indices explicitly. As indicated above, the response-field factors given by (\ref{eq:18}) have two crucial properties affecting the selection of terms that can or cannot contribute to the perturbation series: the Kronecker symbol appearing there identifies the two particles with indices $j_s$ and $j_c$ and thus assigns the same particle to the wave vectors labelled by $c$ and $s$. Furthermore, since the propagator $g_{qp}(t_s,t_c)$ vanishes if $t_s \le t_c$, only such terms can contribute for which $t_s > t_c$.

Applying $r$ one-particle density and $m$ one-particle response-field operators to the free generating functional thus leaves us with the expression
\begin{equation}
\fl
\label{eq:61}
  \prod_{c=r-m+1}^rb_{j_c}(c)\,Z_0[\{j_1,\ldots,j_r\}] 
  =\prod_{c=r-m+1}^r\left(
    \ii\sum_{s=1}^r\left(\vec k_c\cdot\vec k_s\right)g_{qp}(t_s,t_c)
    \delta_{j_cj_s}
  \right)Z_0[\{j_1,\ldots,j_r\}]\;.
\end{equation}
Several aspects are important to note at this point. First, each response-field factor with index $c$ identifies two particles with the indices $j_c$ and $j_{s_c}$. Second, no terms can contribute to any response-field factor which belong to the same times because $g_{qp}(t, t) = 0$. Therefore, only such particles may be identified which are assigned to wave vectors at different times. Fourth, since $g_{qp}(t_1, t_2) = 0$ for $t_2\ge t_1$, any response-field factor implies a time ordering in the sense $t_1 > t_2$.

Beginning with a set of particle indices $\{j_1,\ldots,j_r\}$, we thus proceed as follows to evaluate the terms in (\ref{eq:61}): We first identify as many particle pairs as response fields occur, i.e.\ we identify $m$ pairs of particle indices, taking care that particle pairs must not identify positions with equal times and that a time-ordering is involved in all response-field factors. The remaining $r-m$ particles form a reduced set $\{j\}'$ of particle indices. All possible reduced sets $\{j\}'$ must finally be summed over.

Let us illustrate this procedure with the simple example $r = 4$ and $m = 1$, corresponding to the terms contributing to a two-point density correlator calculated at first-order in the particle interaction. Since the interaction is instantaneous, $t_3 = t_4$, and if the correlator is chosen to be simultaneous, $t_1 = t_2$.

In this case, (\ref{eq:61}) simplifies to the two possible terms
\begin{eqnarray}
  &\ii\left(\vec k_4\cdot\vec k_1\right)g_{qp}(t_1,t_4)\,
  Z_0[\{j_1=j_4,j_2,j_3\}] \;,\nonumber\\
  &\ii\left(\vec k_4\cdot\vec k_2\right)g_{qp}(t_2,t_4)\,
  Z_0[\{j_1,j_2=j_4,j_3\}]\;.
\label{eq:62}
\end{eqnarray}
Other terms would identify particles at the same time and thus return zero.

\subsection{Diagrams}

Combining (\ref{eq:37}) and (\ref{eq:53}) can profitably be represented by diagrams which greatly help constructing and ordering the terms appearing in the generating functional. The essential point of the diagrammatic representation which we are going to construct now is to systematically construct all wave vectors $\vec k_{jk}$ according to (\ref{eq:85}) which enter into the factors $(\mathcal{P}_{jk}+\Delta_{jk})$ appearing in the generating functional.

\begin{figure}[ht]
  \centerline{\includegraphics[height=0.5\hsize]{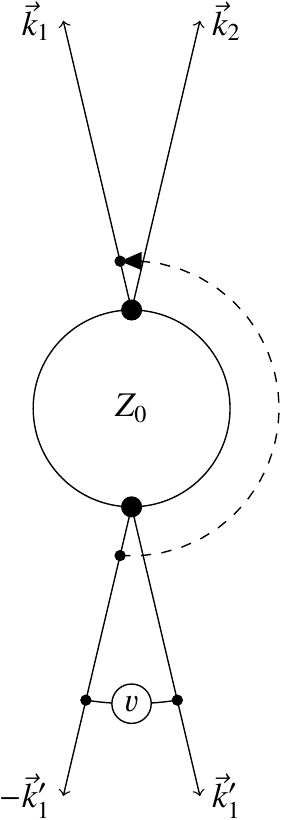}}
\caption{Diagram representing the contribution of the first-order interaction term to a two-point density correlator. The diagram has a multiplicity of $2$ because the wave vectors $\vec k_1$ and $\vec k_2$ are indistinguishable in a homogeneous random field.}
\label{fig:6}
\end{figure}

The diagrams are constructed according to the following rules:

\begin{enumerate}
  \item Mark the free generating functional $Z_0[\tens J, \tens K]$ by a circle. According to (\ref{eq:10}) and (\ref{eq:12}), each one-particle density operator $\hat\rho_{j_s}$ applied to $Z_0$ corresponds to a functional derivative with respect to a component $\vec J_{q_{j_s}}$ of the generator field $\tens J$. According to (\ref{eq:12}) and (\ref{eq:13}), each of these operations adds a wave vector $\vec k_s$ to the shift tensor $\tens L$ that we need to find for each term in a perturbation series.
  
  \emph{Thus, for each of the $r = n+2m$ density operators applied for a perturbation term appearing in an $n$-th order correlator at $m$-th order in the particle interaction, we attach a wave vector to the circle symbolizing $Z_0$, pointing outward.}
  
  \item The times when these density operators act are represented by filled dots on the circumference of the $Z_0$ symbol. Each $\vec k$ vector thus begins at a filled dot. Since each interaction is instantaneous, all internal $\vec k$ vectors need to be pairwise attached to the same time. If the correlator to be calculated is simultaneous, the external $\vec k$ vectors are also attached to the same time.
  
  \emph{Thus, the internal wave vectors representing interactions are pairwise attached to the same points in time. Times are ordered counter-clockwise, with the latest times appearing on top. The external wave vectors appearing in the final correlator are attached to the same time if the correlator is simultaneous.}
  
  \item Between the internal wave vectors corresponding to the density and the response-field operator of an interaction term (\ref{eq:9}), a further factor appears representing the interaction potential $v$. If this potential depends only on the separation between the two particles interacting, it requires the two $\vec k$ vectors of the density and the response-field operator it connects to be equal in magnitude and opposite in sign.
  
  \emph{Internal wave vectors are marked with primes. To include the interaction potential into the diagram, we connect any two internal wave vectors with a circled $v$ which enter into a single interaction term. For translation-invariant potentials $v$, the two internal wave vectors connected by a potential are equal and opposite.}
  
  \item According to (\ref{eq:18}), each response field identifies two particles at different times, i.e.\ it assigns the same particle to two different wave vectors at two different times. Adopting the convention in (\ref{eq:9}), we assign response fields to the negative internal wave vectors appearing in the interaction terms.
  
  \emph{Thus, response fields are represented by dashed circle segments between two different wave vectors attached to different times. Each response field must begin at a negative internal wave vector and must be connected to exactly one other wave vector, internal or external.}
  
  \item Equivalent diagrams can appear multiple times. For example, in the diagram shown in Fig.~\ref{fig:6} representing a contribution to the density power spectrum at first order in the particle interactions, the response-field arrow can end on $\vec k_1$ or $\vec k_2$. In a homogeneous random field, these two wave vectors are equivalent, and the diagram with the response-field arrow attached to the vector $\vec k_2$ corresponds to an identical perturbation term.
  
  \emph{Thus, diagrams are assigned a multiplicity corresponding to the number of equivalent configurations they express.}
\end{enumerate}

Figure~\ref{fig:6} shows the single diagram according to these rules representing the first-order interaction contribution to the second-order density power spectrum.

\begin{figure}[ht]
  \hfill\includegraphics[width=0.83\hsize]{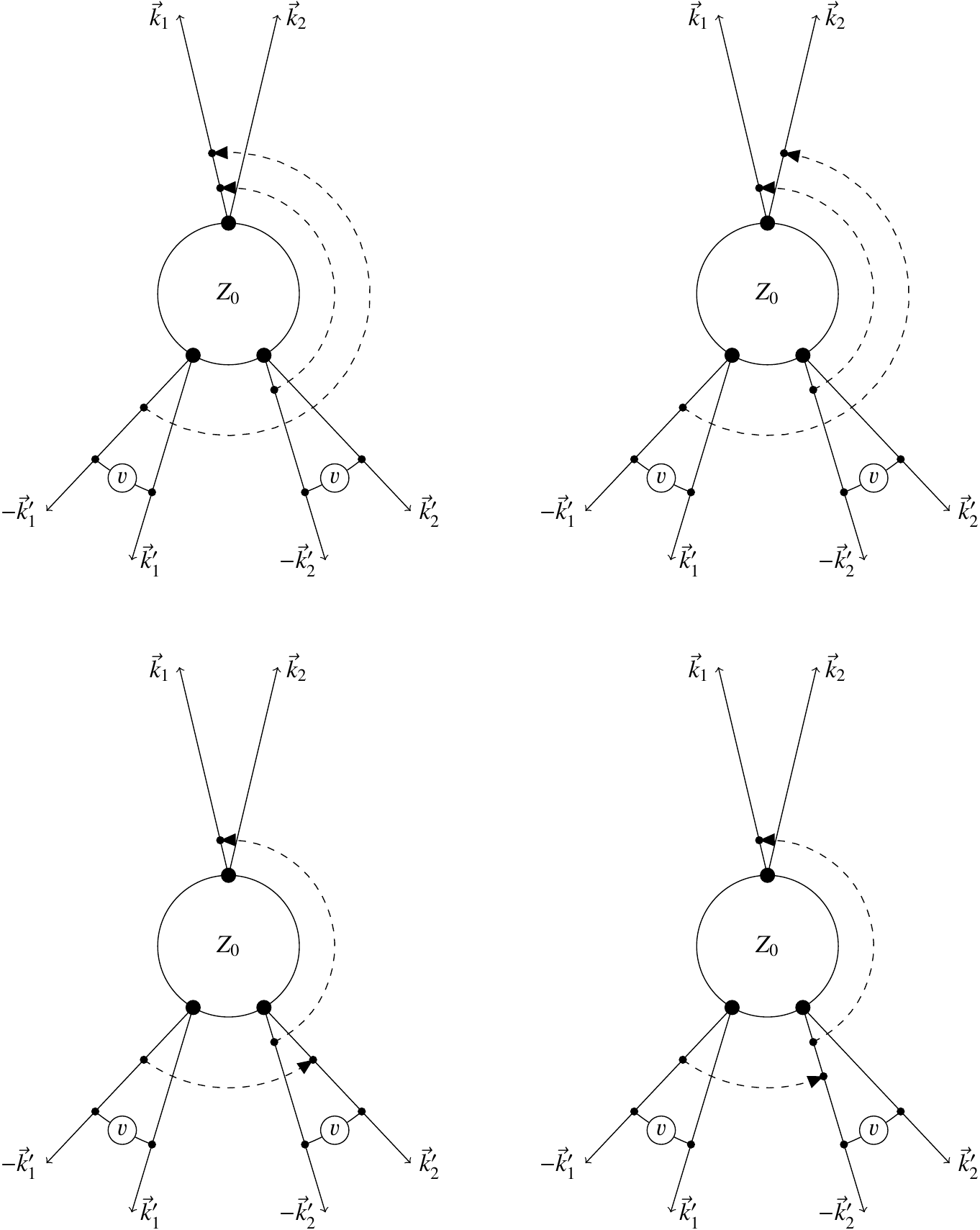}
\caption{Diagrams representing the terms contributing to a two-point correlator at second order in the interaction. Both diagrams in the top row have a multiplicity of $2$ because $\vec k_1$ and $\vec k_2$ can be interchanged in a homogeneous random field. The multiplicity of the diagrams in the bottom row is $4$ because the time order of the two interactions can be exchanged.}
\label{fig:7}
\end{figure}

Diagrams for higher-order interactions or correlators of higher order are now easily constructed. To give an example, we show in Fig.~\ref{fig:7} the four non-equivalent diagrams contributing to a second-order density correlator at second order in the particle interactions.

Position and momentum shift vectors according to (\ref{eq:59}) can now be read off these diagrams as follows: Randomly assign particle indices to the $\vec k$ vectors extending from $Z_0$, thereby assigning the same particle index to such $\vec k$ vectors connected by a dashed line representing a response field. These will be $r-m$ indices in total, for which the numbers $1\ldots r-m$ can be chosen without loss of generality. Any permutation of these indices will result in an equivalent set of shift vectors.

For example, we can use the diagram in Fig.~\ref{fig:6} to assign the particle indices $(1,2,3)$ clock-wise to the $\vec k$ vectors extending from $Z_0$, beginning with $\vec k_1$. This results in the assignment of particles $(1,2,3,1)$ to the wave vectors $(\vec k_1, \vec k_2, \vec k_1', -\vec k_1')$ because $\vec k_1$ and $-\vec k_1'$ are connected by a response field. With (\ref{eq:59}), this implies the position shift vectors
\begin{equation}
  \vec L_{q_1} = -(\vec k_1-\vec k_1')\;,\quad
  \vec L_{q_2} = -\vec k_2\;,\quad
  \vec L_{q_3} = -\vec k_1'
\label{eq:63}
\end{equation}
and the momentum shift vectors
\begin{eqnarray}
  \vec L_{p_1} &= -\left(g_{qp}(t,0)\vec k_1-g_{qp}(t',0)\vec k_1'\right)\;,\quad
  \vec L_{p_2} = -g_{qp}(t,0)\vec k_2\;,
  \nonumber\\
  \vec L_{p_3} &= -g_{qp}(t',0)\vec k_1'\;,
\label{eq:64}
\end{eqnarray}
with $t'$ denoting the time of the interaction and $t>t'$ the time where the correlator is to be evaluated. Any permutation of the particle indices would merely permute the labels on these shift vectors.

These diagrams allow a quick construction of all terms contributing to the generating functional $Z$ for given orders of correlators and of particle interactions. From these diagrams, the shift vectors $\vec L_q$ and $\vec L_p$ as well as the response-field factors can be read off. They can then be inserted into the complete factorization of the generating functional to evaluate  the perturbation terms. The procedures involved can now be implemented in a symbolic computer code.

\section{Summary and conclusions}
\label{sec:5}

In \cite{2016NJPh...18d3020B}, we have developed a kinetic non-equilibrium field theory for cosmic structure formation. The central object of this theory is a free generating functional which describes how an initially correlated ensemble of classical particles propagates in time under Hamiltonian dynamics. Particle interactions are included by an exponential interaction operator whose series expansion suggests a natural perturbative approach. The initial correlation of the particle momenta was shown in \cite{2016NJPh...18d3020B} to be described by a Gaussian in which the momentum-correlation matrix enters as a quadratic form. In \cite{2016NJPh...18d3020B}, we expanded this Gaussian to second order in the quadratic form and showed that, to first order in the particle interactions, the non-linear cosmic density-fluctuation power spectrum known from numerical simulations could be well reproduced. To a large part due to the expansion in the order of the momentum correlations, the notation as well as practical calculations became quite cumbersome.

In this paper, we have substantially simplified the theory by factorising the free generating functional into factors of universal shape, taking the complete hierarchy of momentum correlations into account. Our first main results are thus the expressions (\ref{eq:37}) for the free generating functional and (\ref{eq:38}) for the non-linearly evolving density-fluctuation power spectra $\mathcal{P}$. The factorization suggests the expansion (\ref{eq:53}) of the free generating functional in powers of $\mathcal{P}$, which contain the full hierarchy of momentum correlations. This expansion in terms of power-spectrum factors together with the expansion of the interaction operator suggest a diagrammatic representation of perturbation terms, which we have developed here. 

Our second main result is that the initial momentum correlations of the dark-matter particles, prior to any gravitational interaction, lead to a characteristic deformation of the density-fluctuation power spectrum compared to its initial, linearly evolved shape. For particle momenta aligned with the separation vector between the particles, power is enhanced on moderate and small scales, while any misalignment between these two vectors leads to a substantial transport of power from large to smaller scales.

Our third main result is the development of a diagrammatic approach to the perturbative terms developed from the complete factorization of the generating functional. Taken together, the factorization of the free generating functional and the diagrammatic approach to perturbation theory open a way to have perturbation terms automatically calculated and evaluated by a combination of a symbolic and a numerical computer code, which we are now planning to develop.

\ack

We are grateful for many helpful comments and discussions to J\"urgen Berges, Manfred Salmhofer and Bj\"orn Sch\"afer. This work was supported in part by the German Excellence Initiative and by the Collaborative Research Centre TR 33 ``The Dark Universe'' of the German Science Foundation.

\appendix

\section{Potential correlation function}
\label{sec:A1}

\subsection{Relation to the density-fluctuation power spectrum}

\begin{figure}[ht]
  \hfill\includegraphics[width=0.83\hsize]{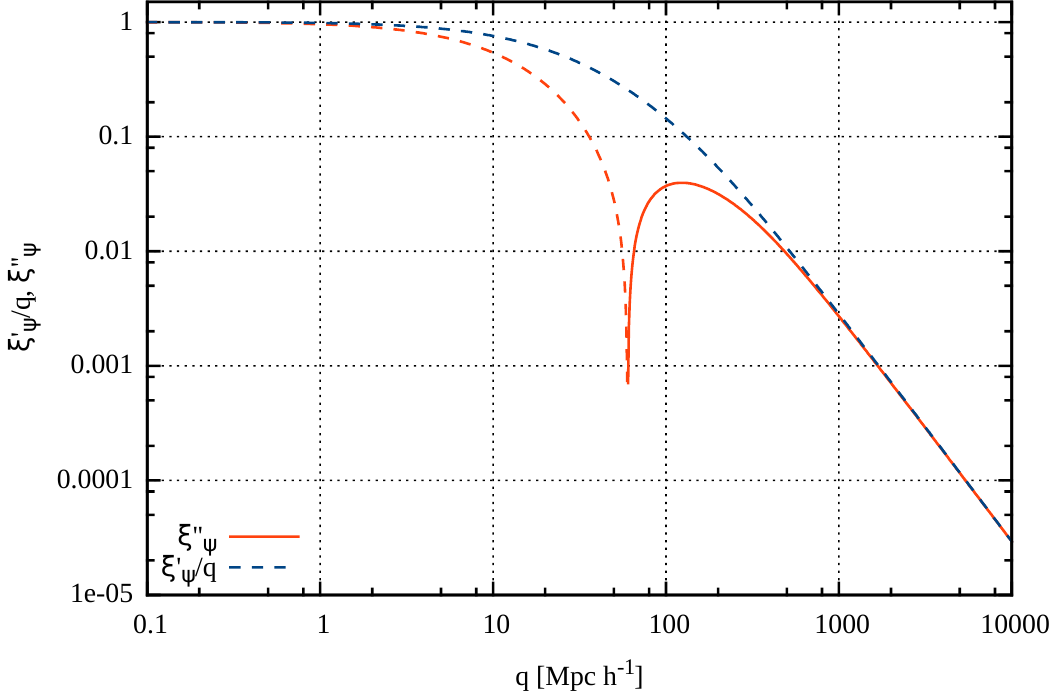}
\caption{First and second derivatives of the potential correlation function $\xi_\psi(q)$, normalized to $\sigma_1^2/3$. Dashed curves and parts thereof indicate negative values.}
\label{fig:9}
\end{figure}

The potential correlation function $\xi_\psi(x)$ introduced in (\ref{eq:29}) is
\begin{equation}
  \xi_\psi(q) = \frac{1}{2\pi^2}
  \int_0^\infty k^2\d k\,P_\psi(k)\,\besselj{0}(kq)\;,
\label{eq:65}
\end{equation}
where $P_\psi$ the power spectrum of the velocity potential. With
\begin{equation}
  \partial_q^\alpha\xi_\psi(q) = \frac{1}{2\pi^2}
  \int_0^\infty k^{2+\alpha}\d k\,P_\psi(k)\,\besselj{0}^{(\alpha)}(kq)
\label{eq:66}
\end{equation}
and using the recursion relations for the spherical Bessel functions and their derivatives, we find
\begin{eqnarray}
\label{eq:67}
  \xi_\psi'(q)  &= -\frac{1}{2\pi^2}
  \int_0^\infty\frac{\d k}{k}P_\delta(k)\,\besselj{1}(kq)\;,\\
  \xi_\psi''(q)-\frac{\xi_\psi'(q)}{q} &=
  \frac{1}{2\pi^2}\int_0^\infty\d k\,P_\delta(k)\,\besselj{2}(kq)\;,
\end{eqnarray}
where the density-fluctuation power spectrum $P_\delta(k) = k^4P_\psi(k)$ was introduced. The asymptotic behaviour for small arguments of the spherical Bessel functions of the first kind ensure that
\begin{equation}
  \lim_{q\to0}\frac{\xi_\psi'(q)}{q} = -\frac{\sigma_1^2}{3} =
  \lim_{q\to0}\xi_\psi''(q)\;.
\label{eq:68}
\end{equation}
The functions $\xi_\psi'(q)/q$ and $\xi_\psi''(q)$ are shown in Fig.~\ref{fig:9}.

\subsection{Approximations for momentum correlations}

In view of later numerical evaluations, it will be advantageous to introduce
\begin{equation}
  a_1(q) := \frac{\xi_\psi'(q)}{q}\;,\quad
  a_2(q) := \xi_\psi''(q)-\frac{\xi_\psi'(q)}{q}\;.
\label{eq:69}
\end{equation}
For the $\Lambda$CDM density-fluctuation power spectrum, it turns out that these two functions $a_1$ and $a_2$ allow simple fits by rational functions,
\begin{eqnarray}
  a_1 &= A_1\left(1+\frac{q}{q_{11}}\right)^{-\alpha_1}\;,\nonumber\\
  a_2 &= A_2q^{\alpha_2}\left(
      1+\left(\frac{q}{q_{21}}\right)^{\beta_2}+
      \left(\frac{q}{q_{22}}\right)^{\gamma_2}
    \right)^{-\delta_2}\;,
\label{eq:70}
\end{eqnarray}
both of which fall like $q^{-2}$ asymptotically for $q\gg q_1$ or $q\gg q_3$. For a $\Lambda$CDM power spectrum according to \cite{1986ApJ...304...15B} with cosmological parameters $\Omega_\mathrm{m0} = 0.3$ and $\Omega_{\Lambda0} = 0.7$, we find the best-fitting parameters listed in Tab.~\ref{tab:1}.

\begin{table}
\caption{Parameters of the rational functions (\ref{eq:70}) fitting the functions $a_1$ and $a_2$ defined in (\ref{eq:69}) for a $\Lambda$CDM power spectrum according to \cite{1986ApJ...304...15B} with cosmological parameters $\Omega_\mathrm{m0} = 0.3$ and $\Omega_{\Lambda0} = 0.7$.}
\label{tab:1}
\begin{center}
\begin{tabular}{|lr|lr|}
\hline
$\ln A_1$ & $-10.0039$ & $\ln A_1$ & $-11.2727$ \\
$q_{11}$ & $61.6918$ & $q_{21}$ & $0.3608$ \\
& & $q_{22}$ & $11.9575$ \\
$\alpha_1$ & $2.0507$ & $\alpha_2$ & $2.04459$ \\
& & $\beta_2$ & $0.5645$ \\
& & $\gamma_2$ & $1.7061$ \\
& & $\delta_2$ & $2.3780$ \\
\hline
\end{tabular}
\end{center}
\end{table}

\begin{figure}[ht]
  \hfill\includegraphics[width=0.83\hsize]{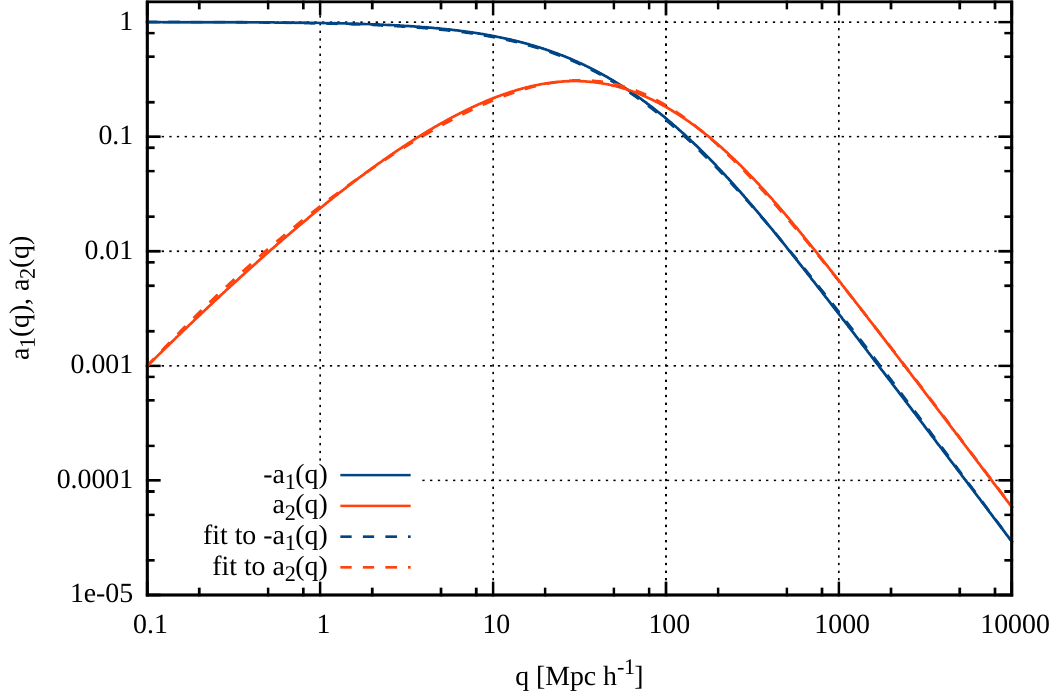}
\caption{The functions $a_{1,2}(q)$, normalized to $\sigma_1^2/3$, are shown here together with the fit functions (\ref{eq:70}) specified by the parameters given in Tab.~\ref{tab:1}.}
\label{fig:10}
\end{figure}

Figure~\ref{fig:10} shows the functions $a_{1,2}(q)$ together with the fits (\ref{eq:70}) specified by the parameters in Tab.~\ref{tab:1}.

\section{Factorization of the generating functional}
\label{sec:A2}

In this Section of the Appendix, we factorize the free generating functional (\ref{eq:23}). We have shown in \cite{2016NJPh...18d3020B} that, due to the scaling of the interaction potential with the inverse mean particle density and the identification of particles with each other by the response field, the calculation of a non-shot noise contribution to an $n$-point correlator at $m$-th order in the particle interactions requires the contribution of order $r = n+2m$ to the free density correlator due to $l = n+m$  particles. Thus, there are $l$ entries in $\tens{L}_{q,p}$ and uncorrelated particles can simply be integrated out from (\ref{eq:23}). Generally, we denote the number of entries in $\tens L_{q,p}$ by $l$ and enumerate them consecutively from $1\ldots l$, which is possible without loss of generality because the enumeration of the particles is unimportant.

\subsection{Introducing relative particle coordinates}

Returning to (\ref{eq:23}), we introduce the coordinate differences with respect to the arbitrarily chosen particle $1$,
\begin{equation}
  \vec q_{j1} := \vec q_j-\vec q_1\quad\forall\quad j=2\ldots l\;.
\label{eq:71}
\end{equation} 
Then, the scalar product of the positions $\tens q$ with the spatial shift tensor $\tens L_q$ can be written as
\begin{equation}
  \left\langle\tens L_q,\tens q\right\rangle =
  \vec L_{q_1}\cdot\vec q_1+
  \vec L_{q_2}\cdot\left(\vec q_{21}+\vec q_1\right)+\ldots =
  \left(\sum_{j=1}^l\vec L_{q_j}\right)\vec q_1+
  \sum_{j=2}^l\vec L_{q_j}\cdot\vec q_{j1}\;.
\label{eq:72}
\end{equation}
Since the correlation matrix $C_{pp}$ depends on coordinate differences only and not on absolute positions, the integration over $\vec q_1$ can be carried out. This results in
\begin{equation}
\label{eq:73}
  Z_0[\tens L,0] = \frac{(2\pi)^3}{V^l}
  \delta_\mathrm{D}\left(\sum_{j=1}^l\vec L_{q_j}\right)
  \prod_{j=2}^l\int_{q_{j1}}
  \exp\left(
    -\frac{1}{2}\tens L_p^\top C_{pp}\tens L_p+
    \ii\sum_{j=2}^l\vec L_{q_j}\cdot\vec q_{j1}
  \right)\;.
\end{equation}
The momentum-correlation matrix depends on the absolute values of all pair-wise coordinate differences
\begin{equation}
  \vec q_{jk} := \vec q_j-\vec q_k
\label{eq:74}
\end{equation}
with
\begin{equation}
  k=1\ldots(l-1)\;,\quad j=(k+1)\ldots l\;,
\label{eq:75}
\end{equation} 
not just on the coordinate differences $\vec q_{j1}$ with respect to particle number $1$. Since there are $l$ particles to consider in total, the number of coordinate differences that $C_{pp}$ depends on is
\begin{equation}
  N_\mathrm{pairs} = \frac{l(l-1)}{2}\;.
\label{eq:76}
\end{equation} 
Of these coordinate differences, $(l-1)$ are taken into account by the $(l-1)$ difference vectors $\vec q_{j1}$. We now extend the integration in (\ref{eq:73}) to \emph{all} these coordinate differences by introducing the remaining difference vectors
\begin{equation}
  \vec q_{ab} = \vec q_a-\vec q_b = \vec q_{a1}-\vec q_{b1}
\label{eq:77}
\end{equation}
with
\begin{equation}
  b = 2\ldots(l-1)\;,\quad a = (b+1)\ldots l\;.
\label{eq:78}
\end{equation} 
These contribute
\begin{equation}
  \frac{l(l-1)}{2}-(l-1) = \frac{(l-1)(l-2)}{2}
\label{eq:79}
\end{equation}
additional, dependent coordinates related by (\ref{eq:77}), which must be ensured by including appropriate delta distributions
\begin{equation}
  \delta_\mathrm{D}\left(\vec q_{ab}-\vec q_{a1}+\vec q_{b1}\right)
\label{eq:80}
\end{equation}
into the integrand, with $a$ and $b$ from (\ref{eq:78}). Thus, the free generating functional turns into
\begin{eqnarray}
\label{eq:81}
  Z_0[\tens L,0] &= \frac{(2\pi)^3}{V^l}
  \delta_\mathrm{D}\left(\sum_{j=1}^l\vec L_{q_j}\right)
  \prod_{j>k}\int_{q_{jk}}
  \exp\left(
    -\frac{1}{2}\tens L_p^\top C_{pp}\tens L_p+
    \ii\sum_{j=2}^l\vec L_{q_j}\cdot\vec q_{j1}
  \right) \nonumber\\ &\cdot
  \prod_{a>b}
  \delta_\mathrm{D}\left(\vec q_{ab}-\vec q_{a1}+\vec q_{b1}\right)\;.
\end{eqnarray}

Replacing all of these delta distributions by their Fourier expansions,
\begin{equation}
  \delta_\mathrm{D}\left(\vec q_{ab}-\vec q_{a1}+\vec q_{b1}\right) =
  \int_{k_{ab}}
  \e^{\ii\vec k'_{ab}\cdot\left(\vec q_{ab}-\vec q_{a1}+\vec q_{b1}\right)}
\label{eq:82}
\end{equation}
with auxiliary wave vectors $\vec k'_{ab}$ conjugate to the coordinate \emph{differences} $\vec q_{ab}$, we arrive at the expression
\begin{eqnarray}
\label{eq:83}
  Z_0[\tens L,0] &= \frac{(2\pi)^3}{V^l}
  \delta_\mathrm{D}\left(\sum_{j=1}^l\vec L_{q_j}\right)
  \prod_{a>b}\int_{k'_{ab}}
  \prod_{j>k}\int_{q_{jk}} \\ &\cdot
  \exp\Bigg(
    -\frac{1}{2}\tens L_p^\top C_{pp}\tens L_p+
    \ii\sum_{j=2}^l\vec L_{q_j}\cdot\vec q_{j1} +
    \ii\sum_{a>b}\vec k'_{ab}\cdot
    \left(\vec q_{ab}-\vec q_{a1}+\vec q_{b1}\right)
  \Bigg) \nonumber
\end{eqnarray}
for the free generating functional. Reordering terms, we can rewrite the phase factor as
\begin{eqnarray}
  &\sum_{j=2}^l\vec L_{q_j}\cdot\vec q_{j1}+
  \sum_{a>b}\vec k'_{ab}\cdot\left(\vec q_{ab}-\vec q_{a1}+\vec q_{b1}\right)
  \nonumber\\ &=
  \sum_{j=2}^l\left(
    \vec L_{q_j}-\sum_{b=2}^{j-1}\vec k'_{jb}+\sum_{a=j+1}^l\vec k'_{aj}
  \right)\cdot\vec q_{j1}+
  \sum_{a>b}\vec k'_{ab}\cdot\vec q_{ab} =:
  \sum_{j>k}\vec k_{jk}\cdot\vec q_{jk}
\label{eq:84}
\end{eqnarray}
with
\begin{equation}
\fl
  \vec k_{jk} := \left\{
  \begin{array}{lll}
    \vec L_{q_j}-\sum_{b=2}^{j-1}\vec k'_{jb}+\sum_{a=j+1}^l\vec k'_{aj}
    & \mbox{for} & \quad k = 1\;,\quad j=2\ldots l \\
    \vec k'_{jk}
    & \mbox{for} & \quad k=2\ldots(l-1)\;,\quad j=(k+1)\dots l
  \end{array}\right.
\label{eq:85}
\end{equation}
and $a,b$ from (\ref{eq:78}).

With these results and the definition of $Q_0$ and $Q_\mathrm{D}$ in (\ref{eq:35}), the generating functional (\ref{eq:83}) factorizes completely in the integrations over the coordinate differences $\vec q_{jk}$,
\begin{equation}
  Z_0[\tens L,0] = \frac{(2\pi)^3}{V^l}
  \delta_\mathrm{D}\left(\sum_{j=1}^l\vec L_{q_j}\right)
  \e^{-(Q_0-Q_\mathrm{D})/2}
  \prod_{a>b}\int_{k_{ab}}
  \prod_{j>k}I_{jk}
\label{eq:86}
\end{equation}
with
\begin{equation}
  I_{jk} := \int_{q_{jk}}\e^
  {-\vec L_{p_j}^\top C_{p_jp_k}\vec L_{p_k}+\ii\vec k_{jk}\cdot\vec q_{jk}}
  \quad (j\ne k)
\label{eq:87}
\end{equation} 
since the correlation matrix $C_{p_jp_k}$ depends on the distances $\vert\vec q_{jk}\vert$ between the particles only. We can thus decompose the generating functional into independent factors for all particle pairs, which are then to be convolved in Fourier space by integrating over all auxiliary wave vectors $\vec k'_{ab}$.

\subsection{Examples}

For two-point density correlations, $l = 2+m$, hence the number of coordinate pairs is
\begin{equation}
  N_\mathrm{pair} = \frac{(m+2)(m+1)}{2}\;,
\label{eq:88}
\end{equation}
of which
\begin{equation}
  \frac{(l-1)(l-2)}{2} = \frac{m(m+1)}{2}
\label{eq:89}
\end{equation}
are dependent.

For first-order perturbation theory of a two-point spectrum, we have $l = 3$, and we need to introduce one auxiliary wave vector $\vec k'_{32}$. According to (\ref{eq:85}), we then have
\begin{equation}
  \vec k_{21} = \vec L_{q_2}+\vec k'_{32}\;,\quad
  \vec k_{31} = \vec L_{q_3}-\vec k'_{32}\;,\quad
  \vec k_{32} = \vec k'_{32}\;,
\label{eq:90}
\end{equation}
and $\vec L_{q_1} = -(\vec L_{q_2}+\vec L_{q_3})$ because of the $\delta$ distribution in (\ref{eq:73}).

For second-order perturbation theory of a two-point spectrum, $l = 4$, and three auxiliary wave vectors $\vec k'_{32}$, $\vec k'_{42}$ and $\vec k'_{43}$ need to be introduced. Then,
\begin{eqnarray}
  \vec k_{21} &= \vec L_{q_2}+\vec k'_{32}+\vec k'_{42}\;,\nonumber\\
  \vec k_{31} &= \vec L_{q_3}-\vec k'_{32}+\vec k'_{43}\;,\nonumber\\
  \vec k_{41} &= \vec L_{q_4}-\vec k'_{42}-\vec k'_{43}\;,\nonumber\\
  \vec k_{32} &= \vec k'_{32}\;,\quad
  \vec k_{42} = \vec k'_{42}\;,\quad
  \vec k_{43} = \vec k'_{43}\;,
\label{eq:91}
\end{eqnarray}
and $\vec L_{q_1} = -(\vec L_{q_2}+\vec L_{q_3}+\vec L_{q_4})$\;.

\subsection{Evaluation of the generic factors}

We now need to evaluate the integrals $I_{jk}$ defined in (\ref{eq:87}), which are all of the type
\begin{equation}
  I_{21} := \int_q\,\e^
  {-\vec L_{p_2}^\top C_{p_2p_1}\vec L_{p_1}+\ii\vec k_{21}\cdot\vec q}\;,
\label{eq:92}
\end{equation}
with an independent wave vector $\vec k_{21}$ representing any of the vectors $\vec k_{jk}$ defined in (\ref{eq:85}).

The momentum-correlation matrix $C_{p_2p_1}$ defined in (\ref{eq:32}) depends on the particle separation $q$ only. Repeating (\ref{eq:32}), we can write $C_{p_2p_1}$ as
\begin{equation}
   C_{p_2p_1} = -\tilde\pi_\parallel\,\xi_\psi''(q)-
   \tilde\pi_\perp\frac{\xi_\psi'(q)}{q}
\label{eq:93}
\end{equation}
with the projectors $\tilde\pi_\parallel$ and $\tilde\pi_\perp$ defined in (\ref{eq:30}).

We now expand the projectors $\tilde\pi_{\parallel,\perp}$ with respect to $\hat q$ into the respective projectors $\pi_{\parallel, \perp}$ with respect to $\vec k_{21}$. Let $\hat k$ be the unit vector in the direction of $\vec k_{21}$, then
\begin{equation}
  \pi_{21}^\parallel = \hat k\otimes\hat k\;,\quad
  \pi_{21}^\perp = \mathcal{I}_3-\pi_{21}^\parallel\;.
\label{eq:94}
\end{equation}
For doing so, we expand the Hessian of the potential-correlation function into the projectors $\pi_{21}^\parallel$ and $\pi_{21}^\perp$,
\begin{equation}
  D^2\xi_\psi(q) =
  \tilde\pi_\parallel\,\xi_\psi''(q)+\tilde\pi_\perp\frac{\xi_\psi'(q)}{q} =
  a_\parallel\pi_{21}^\parallel+a_\perp\pi_{21}^\perp\;,
\label{eq:95}
\end{equation}
multiply this equation by $\pi_{21}^\parallel$ and $\pi_{21}^\perp$ and take the trace of the resulting two equations to find
\begin{eqnarray}
  a_\parallel &=
  \xi_\psi''(q)\tr\tilde\pi_\parallel\pi_{21}^\parallel+
  \frac{\xi_\psi'(q)}{q}\tr\tilde\pi_\perp\pi_{21}^\parallel\;,\nonumber\\
  2a_\perp &=
  \xi_\psi''(q)\tr\tilde\pi_\parallel\pi_{21}^\perp+
  \frac{\xi_\psi'(q)}{q}\tr\tilde\pi_\perp\pi_{21}^\perp\;.
\label{eq:96}
\end{eqnarray}
Introducing the cosine $\mu := \hat q\cdot\hat k$ of the angle between $\vec q$ and $\vec k_{21}$, the traces on the right-hand sides are
\begin{equation}
\label{eq:97}
  \tr\tilde\pi_\parallel\pi_{21}^\parallel = \mu^2\;,\quad
  \tr\tilde\pi_\perp\pi_{21}^\parallel = 1-\mu^2 =
  \tr\tilde\pi_\parallel\pi_{21}^\perp\;,\quad
  \tr\tilde\pi_\perp\pi_{21}^\perp = 1+\mu^2\;.
\end{equation}
We thus have
\begin{eqnarray}
  a_\parallel &= \mu^2\xi_\psi''(q)+(1-\mu^2)\frac{\xi_\psi'(q)}{q}\;,
  \nonumber\\
  2a_\perp &= (1-\mu^2)\xi_\psi''(q)+(1+\mu^2)\frac{\xi_\psi'(q)}{q}\;,
\label{eq:98}
\end{eqnarray}
and the quadratic form in (\ref{eq:92}) turns into
\begin{equation}
\label{eq:99}
  \vec L_{p_2}^\top C_{p_2p_1}\vec L_{p_1} =
  -\vec L_{p_2}^\top\pi_{21}^\parallel\vec L_{p_1}\,a_\parallel-
  \vec L_{p_2}^\top\pi_{21}^\perp\vec L_{p_1}\,a_\perp\;.
\end{equation}
It is now convenient to quantify the remaining projections by $\lambda_{21}^{\parallel,\perp}$ defined by
\begin{equation}
  \vec L_{p_2}^\top\pi_{21}^\parallel\vec L_{p_1} =
  g_{qp}^2(\tau,0)\,k_{21}^2\lambda_{21}^\parallel\;,\quad
  \vec L_{p_2}^\top\pi_{21}^\parallel\vec L_{p_1} =
  g_{qp}^2(\tau,0)\,k_{21}^2\lambda_{21}^\perp\;.
\label{eq:100}
\end{equation}
With these definitions, we arrive at the form
\begin{equation}
  I_{21} = \int_q\,\e^{g_{qp}^2(\tau,0)\,k_{21}^2\left(
    a_\parallel\lambda_{12}^\parallel+a_\perp\lambda_{21}^\perp
  \right)+\ii\vec k_{21}\cdot\vec q}
\label{eq:101}
\end{equation}
for the integral to be solved.

This integral has an intuitive physical meaning. To clarify it, we first split off a delta distribution,
\begin{equation}
  I_{21} = \Delta_{21}+\mathcal{P}_{21}
\label{eq:102}
\end{equation}
with
\begin{equation}
  \mathcal{P}_{21} :=
  \int_q\,\left\{\e^{g_{qp}^2(\tau,0)\,k_{21}^2\left(
    a_\parallel\lambda_{12}^\parallel+a_\perp\lambda_{21}^\perp
  \right)}-1\right\}
  \e^{\ii\vec k_{21}\cdot\vec q}
\label{eq:103}
\end{equation}
and
\begin{equation}
  \Delta_{21} := (2\pi)^3\delta_\mathrm{D}\left(\vec k_{21}\right)\;.
\label{eq:104}
\end{equation}
Due to the large dynamic range of the argument of the exponential in (\ref{eq:103}) and the fast oscillations of the Fourier phase, the integration required to evaluate $\mathcal{P}_{21}$ is numerically difficult in some regions of parameter space. We use Levin collocation \cite{1997JCApM..78..131L, 1996JCApM..67...95L, 1982MComp..38..531L} for a fast and reliable integration scheme.

The meaning of $\mathcal{P}_{21}$ becomes best visible in the limit of early times. Then, the argument of the exponential in (\ref{eq:103}) is small, the exponential can be approximated by a first-order Taylor expansion, and we are left with
\begin{equation}
  \mathcal{P}_{21} \approx
  g_{qp}^2(\tau,0)\,k_{21}^2\int_q\,\left(
    a_\parallel\lambda_{12}^\parallel+a_\perp\lambda_{21}^\perp
  \right)\,\e^{\ii\vec k_{21}\cdot\vec q}\;.
\label{eq:105}
\end{equation}
To evaluate this expression, we define the integrals
\begin{equation}
  I_n^\alpha(k) := 2\pi\int_0^\infty\d q\,q^n\xi_\psi^{(n)}(q)
  \int_{-1}^1\d\mu\,\mu^\alpha\e^{\ii kq\mu}\;,
\label{eq:106}
\end{equation}
where $\xi_\psi^{(n)}(q)$ is the $n$-th derivative of the correlation function $\xi_\psi(q)$. Using
\begin{equation}
  \int_{-1}^1\d\mu\,\mu^\alpha\e^{\ii kq\mu} =
  \frac{1}{(\ii k)^\alpha}\,\partial_q^\alpha\,
  \int_{-1}^1\d\mu\,\e^{\ii kq\mu} =
  \frac{2}{(\ii k)^\alpha}\,\partial_q^\alpha\,\besselj{0}(kq)
\label{eq:107}
\end{equation}
and the recurrence relations for the derivatives of the spherical Bessel functions, we find
\begin{eqnarray}
  I_1^0(k) &= 4\pi\int_0^\infty\d q\,\xi_\psi(q)
    \left(kq\besselj{1}(kq)-\besselj{0}(kq)\right)\;,\nonumber\\
  I_2^0(k) &= 4\pi\int_0^\infty\d q\,\xi_\psi(q)\left[
    \left(2-k^2q^2\right)\besselj{0}(kq)-2kq\besselj{1}(kq)
  \right]\;,\nonumber\\
  I_1^2(k) &= 4\pi\int_0^\infty\d q\,\xi_\psi(q)\left[
    \besselj{0}(kq)+kq\left(1-\frac{4}{k^2q^2}\right)\besselj{1}(kq)
  \right]\;,\nonumber\\
  I_2^2(k) &= 4\pi\int_0^\infty\d q\,\xi_\psi(q)\left[
    \left(2-k^2q^2\right)\besselj{0}(kq)-\frac{4}{kq}\besselj{1}(kq)
  \right]\;.
\label{eq:108}
\end{eqnarray} 
Using these results with $a_\parallel$ and $a_\perp$ from (\ref{eq:98}), we find immediately
\begin{equation}
  \int_qa_\parallel\e^{\ii\vec k\cdot\vec q} =
  I_2^2(k)+I_1^0(k)-I_1^2(k) = -k^2P_\psi(k)
\label{eq:109}
\end{equation}
and
\begin{equation}
  2\int_qa_\perp\e^{\ii\vec k\cdot\vec q} =
  I_2^0(k)-I_2^2(k)+I_1^0(k)+I_1^2(k) = 0\;.
\label{eq:110}
\end{equation}
Therefore, at early times,
\begin{equation}
  \mathcal{P}_{21} \approx -g_{qp}^2(\tau,0)\,\lambda_{21}^\parallel\,
  k_{21}^4\,P_\psi(k_{21}) =
  -g_{qp}^2(\tau,0)\,\lambda_{21}^\parallel\,P_\delta(k_{21})\;.
\label{eq:111}
\end{equation} 
This is the density-fluctuation power spectrum, linearly evolved with $g_{qp}^2(\tau,0)$, times $-\lambda_{21}^\parallel$. As we shall see in the main text, $\lambda_{21}^\parallel = -1$ in the most straightforward applications. This shows that the function $\mathcal{P}_{21}$ generalises the linearly evolved density-fluctuation power spectrum and thus that $\mathcal{P}_{21}$ is a non-linearly time-evolving density-fluctuation power spectrum. Its Fourier transform is the corresponding generalization of the spatial density correlation function.

\bibliographystyle{iopart-num}
\bibliography{main}

\end{document}